\documentclass[11pt]{article}
\usepackage{latexsym}
\usepackage{amssymb}
\usepackage{amsmath}
\usepackage{graphicx}
\usepackage{colortbl}
\usepackage{tikz}
\usepackage[square,comma]{natbib}

\begin{document}

\def\Nset{\mathbb{N}}
\def\Ascr{\mathcal{A}}
\def\Bscr{\mathcal{B}}
\def\Cscr{\mathcal{C}}
\def\Dscr{\mathcal{D}}
\def\Escr{\mathcal{E}}
\def\Fscr{\mathcal{F}}
\def\Hscr{\mathcal{H}}
\def\Iscr{\mathcal{I}}
\def\Mscr{\mathcal{M}}
\def\Nscr{\mathcal{N}}
\def\Pscr{\mathcal{P}}
\def\Qscr{\mathcal{Q}}
\def\Rscr{\mathcal{R}}
\def\Sscr{\mathcal{S}}
\def\Wscr{\mathcal{W}}
\def\Xscr{\mathcal{X}}
\def\cupp{\stackrel{.}{\cup}}
\def\bold{\bf\boldmath}

\newcommand{\rouge}[1]{\textcolor{red}{\tt \footnotesize #1}}
\newcommand{\boldheader}[1]{\smallskip\noindent{\bold #1:}\quad}
\newcommand{\PP}{\mbox{\slshape P}}
\newcommand{\NP}{\mbox{\slshape NP}}
\newcommand{\opt}{\mbox{\scriptsize\rm OPT}}
\newcommand{\ec}{\mbox{\scriptsize\rm OPT}_{\small\rm 2EC}}
\newcommand{\lp}{\mbox{\scriptsize\rm LP}}
\newcommand{\inn}{\mbox{\rm in}}
\newcommand{\deff}{\mbox{\rm sur}}
\newcommand{\MAXSNP}{\mbox{\slshape MAXSNP}}
\newtheorem{theorem}{Theorem}
\newtheorem{lemma}[theorem]{Lemma}
\newtheorem{corollary}[theorem]{Corollary}
\newtheorem{proposition}[theorem]{Proposition}
\newtheorem{definition}[theorem]{Definition}
\def\prove{\par \noindent \hbox{\bf Proof:}\quad}
\def\endproof{\eol \rightline{$\Box$} \par}
\renewcommand{\endproof}{\hspace*{\fill} {\boldmath $\Box$} \par \vskip0.5em}
\newcommand{\mathendproof}{\vskip-1.8em\hspace*{\fill} {\boldmath $\Box$} \par \vskip1.8em}
\def\cupp{\stackrel{.}{\cup}}

\definecolor{orange}{rgb}{1,0.9,0}
\definecolor{violet}{rgb}{0.8,0,1}
\definecolor{darkgreen}{rgb}{0,0.5,0}
\definecolor{grey}{rgb}{0.75,0.75,0.75}

\title {
\vspace*{-1.8cm}
{\huge Shorter Tours by Nicer Ears:} \\[3mm]
\Large{
7/5-approximation for graphic TSP, \\3/2 for the path version,  \\and 4/3 for two-edge-connected subgraphs\\}
}
\author{Andr\'as Seb\H{o}\footnote{CNRS, UJF, Grenoble-INP,  Laboratoire G-SCOP.
Supported by
the TEOMATRO grant ANR-10-BLAN 0207 ``New Trends in Matroids: Base Polytopes, Structure, Algorithms and Interactions''.}
\and Jens Vygen\footnote{Research Institute for Discrete Mathematics, University of Bonn.
This work was done while visiting Grenoble, Laboratoire G-SCOP. Support of Universit\'e Joseph Fourier is gratefully acknowledged.}}

\begingroup
\makeatletter
\let\@fnsymbol\@arabic
\maketitle
\endgroup

\begin{abstract}
We prove new results for approximating the graphic TSP and some related problems.
We obtain polynomial-time algorithms with improved approximation guarantees.

For the graphic TSP itself, we improve the approximation ratio
to $7/5$.
For a generalization, the connected-$T$-join problem, we
obtain the first nontrivial approximation algorithm, with ratio $3/2$.
This contains the graphic $s$-$t$-path-TSP as a special case.
Our improved approximation guarantee for finding a smallest $2$-edge-connected spanning subgraph is $4/3$.

The key new ingredient of all our algorithms is a special kind of ear-decomposition
optimized using forest representations of hypergraphs.
The same methods also provide the lower bounds (arising from LP relaxations)
that we use to deduce the approximation ratios.

\medskip\noindent
\noindent{{\bf keywords:} traveling salesman problem, graphic TSP, $2$-edge-connected subgraph,
$T$-join, ear-decomposition, matroid intersection, forest representation, matching.}
\end{abstract}

\section{Introduction}

The traveling salesman problem is one of the most famous and notoriously hard combinatorial optimization problems
(\cite{Coo12}).
For 35 years, the best known approximation algorithm for the metric TSP, due to \cite{Chr76}, could not
be improved. This algorithm computes a solution of length at most $\frac{3}{2}$
times the linear programming lower bound (\cite{Wol80}).
It is conjectured that a tour of length at most $\frac{4}{3}$ times
the value of the subtour relaxation always exists:  this is the ratio of the worst known examples.
In these examples the length function on pairs of vertices is
the minimum number of edges of a path between the vertices in an underlying graph.
This natural, purely graph-theoretical special case received much attention recently, and is also the subject of the present work.

\smallskip
\boldheader{Notation and Terminology}
All graphs in this paper are undirected. They can have parallel edges but no loops.
For a graph $G$ we denote by $V(G)$ and $E(G)$ its sets of vertices and edges, respectively.
For $X\subseteq V(G)$ we write $\delta(X)$ for the set of edges with exactly one endpoint in $X$.
We denote by $G[X]$ the subgraph induced by $X$.
An {\em induced matching} in $G$ is the edge-set of an induced subgraph in which all
vertices have degree $1$.
By the {\em components} of $G$ we mean the vertex sets of the maximal connected subgraphs
(so the components form a partition of $V(G)$).
By $2G$ we denote the graph arising from $G$ by doubling all its edges, and a {\em multi-subgraph} of $G$ is a subgraph of $2G$.

If $G$ is a graph and $T\subseteq V(G)$ with $|T|$ even, then  a {\em $T$-join} in $G$ is a set $F\subseteq E(G)$
such that $T=\{v\in V(G): \hbox{$|\delta(v)\cap F|$ is odd}\}.$
The minimum cardinality of a $T$-join in $G$ is denoted by $\tau(G,T)$.
\cite{Edm65} showed how to reduce the minimum (in fact, minimum weight) $T$-join problem
to weighted matching, and thus it can be solved in $O(|V(G)|^3)$ time (\cite{Gab73}).

\vspace{-4pt}
\begin{definition}
A {\em connected-$T$-join of $G$} is a $T$-join $F$ in $2G$ such that $(V(G),F)$ is connected.
If $T=\emptyset$, $F$ will be called  a {\em tour}.
The minimum cardinality of a connected-$T$-join of $G$ is denoted by $\opt(G,T)$,
and the minimum cardinality of a tour by $\opt(G)=\opt(G,\emptyset)$.
\end{definition}
\vspace{-4pt}

The metric closure of a connected graph $G$ is the pair $(\bar G,\bar c)$, where $\bar G$
is the complete graph with $V(\bar G)=V(G)$, and $\bar c(\{v,w\})$ is the minimum number
of edges in a $v$-$w$-path in $G$.

\medskip
\boldheader{Problems}
The {\em graphic TSP} can be described as follows. Given a connected graph $G$, find
\vspace{-6pt}
\begin{itemize}
\addtolength{\itemsep}{-8pt}
\item[-]a shortest Hamiltonian circuit in the metric closure of $G$; or
\item[-]a minimum length closed walk in $2G$ that visits every vertex at least once; or
\item[-]a minimum cardinality connected-$\emptyset$-join of $G$.
\end{itemize}
\vspace{-6pt}

It is easy to see and well-known that these formulations are equivalent;
this is the unweighted special case of the ``graphical TSP'' (see \cite{CFN85}).

We also consider two related problems.
In the {\em connected-$T$-join problem}, the input is a connected graph $G$ and a set
$T\subseteq V(G)$ of even cardinality, and we look for a minimum cardinality connected-$T$-join of $G$.
The case $|T|=2$, say $T=\{s,t\}$, has also been studied and was called the {\em graphic $s$-$t$-path TSP}.
(By ``Euler's theorem'' a subset of $E(2G)$ is a connected-$\{s,t\}$-join if and only if
its edges can be ordered to form a walk from $s$ to $t$
that visits every vertex at least once.)

Note that more than two copies of an edge are never useful.
However, the variants of the above problems that do not allow doubling edges
have no approximation algorithms unless $\PP=\NP$.
To see this, note that in a 3-regular graph any tour without doubled edges is a Hamiltonian circuit,
and the problem of deciding whether a given 3-regular graph is Hamiltonian is $\NP$-complete
(\cite{GarJT76}).

A relaxation of the graphic TSP is the {\em 2-edge-connected subgraph problem}.
Given a connected graph $G$, we look for a
2-edge-connected spanning multi-subgraph with minimum number of edges.
We denote this minimum by $\ec(G)$.
A solution $F$ will of course contain two copies of each bridge,
and may at first contain parallel copies of other edges too.
However, the latter can always be avoided:
if an edge $e$ is not a bridge but has two copies, either the second copy can be deleted from $F$,
or the two copies form a cut in $F$ and, since $e$ is not a bridge in $G$,
there is another edge $f$ between the two sides of this cut; the second copy of $e$ can then be replaced by $f$.
Hence an equivalent formulation asks for
a 2-edge-connected spanning subgraph, called {\em 2ECSS},
with minimum number of edges,
of a given 2-edge-connected graph $G$.
Note that any tour in a 2-edge-connected graph $G$ gives rise to a 2ECSS of $G$ with
at most the same number of edges.

\smallskip
\boldheader{Previous Results}
All the above problems are $\NP$-hard because the 2-edge-connected subgraphs of $G$ with
$|V(G)|$ edges are precisely the Hamiltonian circuits.
A {\em $\rho$-approximation algorithm} is a polynomial-time algorithm that always computes a solution
of value at most $\rho$ times the optimum.
For all our problems, a $2$-approximation algorithm is trivial by taking a spanning tree
and doubling all its edges (for TSP or 2ECSS) or some of its edges (for connected-$T$-joins).

\medskip

For the  TSP with arbitrary metric weights (of which the graphic TSP is a proper special case), \cite{Chr76}
described a $\frac{3}{2}$-approximation algorithm. No improvement on this has been found for 35 years,
but recently there has been some progress for the graphic TSP:

A first breakthrough improving on the $\frac{3}{2}$ (by a very small amount) for a
difficult subproblem appeared in \cite{GLS05}; they considered $3$-connected cubic graphs.
This result has been improved to $\frac{4}{3}$ and generalized to all cubic graphs  by \cite{Boyd11},
who also survey other previous work on special cases.
However, for general graphs there has not been any progress until 2011:

\cite{GhaSS11} gave a $(\frac{3}{2}-\epsilon)$-approximation for a tiny $\epsilon>0$,
using a sophisticated probabilistic analysis.
\cite{MomS11} obtained a 1.461-approximation by a simple and clever polyhedral idea,
which easily yields the ratio $\frac{4}{3}$ for cubic (actually subcubic) graphs, and will also be an important tool in the sequel.
\cite{Muc12} refined their analysis and obtained an approximation ratio of $\frac{13}{9}\approx 1.444$.

The graphic TSP was shown to be $\MAXSNP$-hard by \cite{PapY93}.

\medskip

Several of the above articles apply their method to the graphic $s$-$t$-path TSP as well,
but we found no mention of the connected-$T$-join problem.
However, we note that the natural adaptation
of Christofides' [\citeyear{Chr76}] idea provides a $\frac{5}{3}$-approxima\-tion algorithm for minimum weight connected-$T$-joins
for any non-negative weight function $c$ on $E(G)$.
This was noted for the special case $|T|=2$ by \cite{Hoo91}, but works in general as follows.
Let $F$ be the edge set of a minimum weight spanning tree, and $T'$ such that $F$ is a $(T\triangle T')$-join.
Let $J'$ be a minimum weight $T'$-join.
Then the disjoint union $F\cupp J'$ (taking edges appearing in both sets twice) is a connected-$T$-join, and its cost
is at most $\frac{5}{3}$ times the optimum.
To see this, note that $c(F)$ is at the most the optimum.
We now show that $c(J')\le \frac{2}{3}c(J)$, where
$J$ is a minimum weight connected-$T$-join.
Indeed, $F\cupp J$ is a $T'$-join, and can be partitioned into three $T'$-joins:
$(V(G),F)$ is connected and thus contains a $T'$-join $J_1$,
$(V(G),J)$ is connected and thus contains a $T'$-join $J_2$,
and $J_3:=(F\setminus J_1) \stackrel{.}{\cup} (J\setminus J_2)$ is a $T'$-join.
We conclude that
$3c(J') \le c(J_1)+c(J_2)+c(J_3) = c(F)+c(J)\le 2c(J)$.

\cite{AKS12} improved on Christofides' algorithm for the $s$-$t$-path version and obtained an approximation ratio of $1.619$.
They also obtained a $1.578$-approximation algorithm for the graphic case (i.e., the connected-$\{s,t\}$-join problem).

\medskip

For the 2ECSS problem, \cite{KhuV94} gave a $\frac{3}{2}$-approximation algorithm,
and \cite{CheSS01} improved the approximation ratio to $\frac{17}{12}$.
Better approximation ratios have been claimed, but
to the best of our knowledge, no correct proof has been published.

\smallskip
\boldheader{Our results and methods}
We describe polynomial-time algorithms with approximation ratio $\frac{7}{5}$ for the graphic TSP,
$\frac{3}{2}$ for the general connected-$T$-join problem (including graphic $s$-$t$-path-TSP),
and $\frac{4}{3}$ for the 2ECSS problem.

\medskip

The classical work of \cite{Chr76} is still present:
the roles of the edges in our work can most of the time
be separated to working for ``connectivity'' or ``parity''.
We begin by constructing an appropriate ear-decomposition, using a result of \cite{Fra93}
in a similar way as \cite{CheSS01}.
Ear-decompositions can then be combined in a natural way with an ingenious lemma
of \cite{MomS11},
which corrects the parity not only by adding but also by deleting some edges,
without destroying connectivity.
This fits together with ear-decompositions surprisingly well.
However, this is not always good enough.
It turns out that short and ``pendant'' ears need special care.
We can make all short ears pendant (Section~\ref{scear})
and optimize them in order to need a minimum number of additional edges for connectivity (Section~\ref{sec:earmuff}).
This subtask, which we call earmuff maximization, is related to
matroid intersection and forest representations of hypergraphs.
We use our earmuff theorem and the corresponding lower bound (Section~\ref{sclb}) for
all three problems that we study. We present our algorithms in Section~\ref{scalg}.

\medskip

Let us overview the four main assertions that are animating all the rest of the paper:
a key result that will be used as a first construction for our three approximation results is
that  {\em a connected-$T$-join of cardinality at most $\frac{3}{2}\opt(G,T) +\frac{1}{2}\varphi - \pi$
(and at most $\frac{3}{2}\ec(G) - \pi\le\frac{3}{2}\opt(G) - \pi$ if $T=\emptyset$)
can be constructed in polynomial time} (Theorem~\ref{connTjoinbyearmuff}),
where $\varphi$ and $\pi$ are ``the number of even and the number of pendant ears in a suitable ear-decomposition''.
We postpone the precise details until Subsection~\ref{sub:nicer},
where the main optimization problem we have to solve is also explained.
Section~\ref{sec:earmuff} is technically solving this optimization problem.
The solution is used in Theorem~\ref{connTjoinbyearmuff} and in the lower bounds proving its quality.
In the particular case $T=\emptyset$ this construction provides a tour, which can also be used for a 2ECSS.

Then for our three different approximation algorithms we have three different second constructions
for the case when $\pi$ is ``small''.
A simple inductive construction with respect to the ear-decomposition (Propositions~\ref{earinduction} and \ref{prop:connected})
provides a  connected-$T$-join of cardinality at most $\frac{3}{2}\opt(G,T) - \frac{1}{2}\varphi + \pi$.
We see that the smaller of the two connected-$T$-joins has cardinality at most $\frac{3}{2}\opt(G,T)$ (Theorem~\ref{thm:connTjoin}).

If $T=\emptyset$,
our second construction applies the lemma of \cite{MomS11} to our ear-decomposition,
obtaining the bound  $\frac{4}{3}\opt(G) + \frac{2}{3}\pi$ (Lemma~\ref{swedishcase}).
Therefore the worst ratio is given by $\pi=\frac{1}{10}\opt(G)$,
when both constructions  guarantee $\frac{7}{5}\opt(G)$ (Theorem~\ref{thm:tsp}).
We could use this bound for 2ECSS as well, but
here a simple induction with respect to the number of ears obeys the stronger bound
$\frac{5}{4}\ec(G) + \frac{1}{2}\pi$, and so $\pi=\frac{1}{6}\ec(G)$ provides
the worst ratio of $\frac{4}{3}\ec(G)$ (Theorem~\ref{thm:2ECSS}).

\medskip
\boldheader{Preliminaries}
The natural LP relaxation of the 2ECSS problem is the following:
$$\lp(G) \ := \ \min\left\{ x(E(G)) : x\in\mathbb{R}_{\ge 0}^{E(G)}\!, \
x(\delta(W))\ge 2 \mbox{ for all } \emptyset\not=W\subset V(G) \right\},$$
where we abbreviate $x(S):=\sum_{e\in S}x_e$ as usual.
We can give lower bounds by providing dual solutions to this LP.
Obviously we have:
\begin{proposition}
\label{proplp}
For every connected graph $G$:
$$\opt(G) \ \ge \ \ec(G) \ \ge \ \lp(G) \ \ge \ |V(G)|.$$
\mathendproof
\end{proposition}

For the general connected-$T$-join problem $\lp(G)$ is  not a  valid lower bound; we need a more general setting.
For a partition $\Wscr$ of $V(G)$ we introduce the notation
$$\delta(\Wscr) \ := \ \bigcup_{W\in\Wscr} \delta (W),$$
that is, $\delta(\Wscr)$ is the set of edges that have their two endpoints in different classes of $\Wscr$.

Let $G$ be a connected graph, and $T\subseteq V(G)$ with $|T|$ even.
The following seems to take  naturally an analogous role to $\lp(G)$ for connected-$T$-joins:
\begin{eqnarray*}
\lp(G,T) &\! := \!& \min \Bigl\{ x(E(G)) : x\in\mathbb{R}_{\ge 0}^{E(G)}\!, \
 x(\delta(W)) \ge 2 \mbox{ for all } \emptyset\not=W\subset V(G) \mbox{ with } |W\cap T| \hbox{ even},\\
&& \hspace{4.6cm} x(\delta(\Wscr)) \ge |\Wscr| - 1 \mbox{ for all partitions $\Wscr$ of $V(G)$} \Bigr\} .
\end{eqnarray*}
Note that $\lp(G,\emptyset)=\lp(G)$. We obviously have as well:
\begin{proposition}
\label{proplpT}
For every connected graph $G$ and $T\subseteq V(G)$ with $|T|$ even:
$$\opt(G,T) \ \ge \ \lp(G, T ) \ \ge \ |V(G)|-1.$$
\mathendproof
\end{proposition}
The bound can be tight as every spanning tree is a connected-$T$-join,
where $T$ is the set of its odd degree vertices.
Surprisingly, in our lower bounds we will be satisfied by the relaxation of $\lp(G,T)$
in which ``$|W\cap T|$~even'' is replaced by ``$W\cap T=\emptyset$''.

\smallskip
As a last preliminary remark we note that in all our problems, we can restrict our attention to $2$-vertex-connected graphs
because we can consider the blocks (i.e., the maximal 2-vertex-connected subgraphs) separately:
\begin{proposition}
\label{reduction2connected}
Let $G_1$ and $G_2$ be two connected graphs with $V(G_1)\cap V(G_2)=\{v\}$.
Let $G:=(V(G_1)\cup V(G_2),E(G_1)\cup E(G_2))$, and let
$T\subseteq V(G)$, $|T|$ even.
Let $T_i$ be the even set among $(T\cap V(G_i))\setminus\{v\}$ and $(T\cap V(G_i))\cup\{v\}$ $(i=1,2)$.
Then
$\opt(G,T)=\opt(G_1,T_1)+\opt(G_2,T_2)$,
$\ec(G)=\ec(G_1)+\ec(G_2)$,
and $\lp(G,T)=\lp(G,T_1)+\lp(G,T_2)$.
In particular, any approximation guarantee or integrality ratio valid for $(G_1,T_1)$ and $(G_2,T_2)$
is valid for $(G,T)$.
\end{proposition}

\prove
The connected-$T$-joins of $G$ are precisely the unions of a connected-$T_1$-join of $G_1$
and a connected-$T_2$-join of $G_2$.
The same holds for 2ECSS.
We finally show $\lp(G,T)=\lp(G_1,T_1)+\lp(G_2,T_2)$.
For the inequality ``$\ge$'', observe that any feasible solution of $\lp(G,T)$ splits
into feasible solutions of $\lp(G_1,T_1)$ and $\lp(G_2,T_2)$.
The reverse inequality follows from combining feasible dual solutions
of $\lp(G_1,T_1)$ and $\lp(G_2,T_2)$
to a feasible dual solution of $\lp(G,T)$.
\endproof

\section{Ear-Decompositions \label{scear}}

An {\em ear-decomposition} is a sequence $P_0,P_1,\ldots,P_k$, where
$P_0$ is a graph consisting of only one vertex (and no edge), and for each $i\in\{1,\ldots,k\}$ we have:
\vspace{-5pt}
\begin{enumerate}
\addtolength{\itemsep}{-5pt}
\item[{\rm (a)}] $P_i$ is a circuit sharing exactly one vertex with $V(P_0)\cup\cdots\cup V(P_{i-1})$, or
\item[{\rm (b)}] $P_i$ is a path sharing exactly its two different endpoints with $V(P_0)\cup\cdots\cup V(P_{i-1})$.
\end{enumerate}
\vspace{-3pt}
$P_1,\ldots,P_k$ are called {\em ears}.
$P_i$ is a {\em closed ear} if it is a circuit and an {\em open ear} if it is a path.
A vertex in $V(P_i)\cap (V(P_0)\cup\cdots\cup V(P_{i-1}))$ is called an {\em endpoint} of $P_i$,
even if $P_i$ is closed.  An ear has one or two endpoints; its other vertices will be called {\em internal} vertices.
The set of internal vertices of an ear $Q$ will be denoted by $\inn(Q)$.
We always have $|\inn(Q)|=|E(Q)|-1$, while  $|V(Q)|$ is  $|E(Q)|+1$ or $|E(Q)|$
depending on whether $Q$ is an open or closed ear.
If $P$ and $Q$ are ears and $q\in\inn(Q)$ is an endpoint of $P$,
then we say that $P$ is {\em attached} to $Q$ (at $q$).

$P_0,P_1,\ldots,P_k$ is called an ear-decomposition {\em of} the graph
$P_0+P_1+\cdots+P_k:=(V(P_0)\cup\cdots\cup V(P_k),E(P_1)\cup\cdots\cup E(P_k))$.
It is called {\em open} if all ears except $P_1$ are open.

A graph has an ear-decomposition if and only if it is 2-edge-connected.
A graph has an open ear-decomposition if and only if it is $2$-vertex-connected.
The number of ears in any ear-decomposition of $G$ is $|E(G)|-|V(G)|+1$.
These definitions and statements are due to \cite{Whi32}.

\smallskip
We call  $|E(P)|$ the {\em length}  of a path or of an ear $P$.
An {\em $l$-path} is a path of length $l$, and an {\em $l$-ear} is an ear of length $l$;
an $l$-ear for $l>1$ is said to be {\em nontrivial}.
Minimizing the number of nontrivial ears is equivalent to the 2ECSS problem
because deleting 1-ears maintains 2-edge-connectivity.

\smallskip
Given an ear-decomposition, we call an ear {\em pendant} if it is nontrivial and
there is no nontrivial ear attached to it.

\subsection{Even and short ears}

For an ear $P$ let $\varphi(P)=1$ if $|E(P)|$ is even, and $\varphi(P)=0$ if it is odd.  For a 2-edge-connected graph $G$,
$\varphi(G)$ denotes the minimum number of even ears in an ear-decomposition of $G$, that is,
the minimum of $\sum_{i=1}^k \varphi(P_i)$ over all ear-decompositions of $G$.
This parameter was introduced by \cite{Fra93}, who proved that this minimum can be computed in polynomial time.

Another kind of ears that plays a particular role is $2$-ears and $3$-ears.
We will call these {\em short} ears.
Unlike the number of even ears, we do not know how to minimize the number of short ears efficiently.
However, they can be useful in other ways (cf.\ Section~\ref{sec:earmuff}).
All short ears occurring in this paper will be open, except possibly for the first ear.

Recursion (induction) with respect to new ears is not an optimal way of constructing small $T$-joins
(connected or not) or tours, but it allows to deduce simple upper bounds that depend only
on the graph and hold for all $T$.

Let $G$ be a $2$-edge-connected graph with an ear-decomposition,  $T\subseteq V(G)$, $|T|$ even, and $P$ a pendant ear.
Then $P$ is subdivided into subpaths by the vertices of $\inn(P)\cap T$.
Let us color these subpaths blue and red alternatingly.
To obtain a $T$-join in $G$, we could take the edges of the red subpaths and add them to
an $S$-join (where we define $S$ appropriately) in the subgraph induced by $V(G)\setminus\inn(P)$.
For a connected-$T$-join in $G$, we can take $E(P)$, double the edges of the red subpaths, and proceed as before.
In this case we can in addition delete one pair of parallel edges if there is one.

This yields the following bounds.

We will write  $\gamma(P)=1$ if $P$ is  short and $\inn(P)\cap T=\emptyset$, and $\gamma(P)=0$ otherwise.

\begin{lemma}\label{earinduction}
Let $G$ be a $2$-edge-connected graph with an ear-decomposition,
and $T\subseteq V(G)$, $|T|$ even.
Let $P$ be a pendant ear.
Then there exist $F,F'\subseteq E(P)$ and $S,S'\subseteq V(G)\setminus\inn(P)$ such that:
\vspace{-15pt}
\begin{itemize}
\addtolength{\itemsep}{-3pt}
\item[{\rm (a)}]  $|F|\le \frac{1}{2}|\inn(P)|+ \frac{1}{2}\varphi(P)$,
and $F\cup J$ is a $T$-join in $G$ for every $S$-join $J$ in $G-\inn(P)$.
\item[{\rm (b)}]  $|F'|\le \frac{3}{2}|\inn(P)|+\frac{1}{2}\varphi(P)+\gamma(P)-1$,
and $F'\cup J'$ is a connected-$T$-join of $G$ for every connected-$S'$-join $J'$ of $G-\inn(P)$.
\end{itemize}
\vspace{-3pt}
Such sets $F$ and $F'$  can be computed in $O(|\inn(P)|)$ time.
\end{lemma}

\prove
The vertices of $\inn(P)\cap T$ subdivide $P$ into subpaths, alternatingly colored red and blue.
Let $E_R$ and $E_B$ denote the set of edges of red and blue subpaths, respectively; w.l.o.g., $|E_R|\le|E_B|$.
Let $T_R$ and $T_B$ be the set of vertices having odd degree in $(V(P),E_R)$ and $(V(P),E_B)$,
respectively. Note that $\{E_R,E_B\}$ is a partition of $E(P)$, and
$T_R\cap\inn(P)=T_B\cap\inn(P)=T\cap\inn(P)$.

Let $S:=T\Delta T_R$ and $F:=E_R$.
Then $F$ and $S$ satisfy the claims in (a) because
 $|F|\le \lfloor \frac{1}{2} |E(P)|\rfloor=\frac{1}{2}(|\inn(P) |+\varphi(P))$.

For (b) let $S':=T\Delta T_B$. We distinguish two cases.
If $E_R=\emptyset$, then let $F':=E_B=E(P)$.
Then
$\textstyle
|F'| \ = \ |E(P)| \ = \ |\inn(P)|+1 \ \le \ \frac{3}{2}|\inn(P)|+\frac{1}{2}\varphi(P) + \gamma(P) - 1.$

If $E_R\not=\emptyset$, then let $F'$ result from $E(P)$ by
doubling the edges of $E_R$ and then removing one arbitrary pair of parallel edges.
Using (a) we have
$$\textstyle
|F'| \ = \ |E(P)| + |E_R| - 2 \ = \ |\inn(P)| + 1 + |F| - 2
\ \le \ \frac{3}{2}|\inn(P)|+\frac{1}{2}\varphi(P) - 1.$$
\mathendproof

\begin{proposition}[\cite{Fra93}]
\label{prop:frank}
Let $G$ be a $2$-edge-connected graph, and $T\subseteq V(G)$, $|T|$ even. Then
$$\textstyle
\tau(G,T) \ \le \ \frac{1}{2}(|V(G)|+\varphi(G)-1).$$
\end{proposition}

\prove   Let $P_0,\ldots,P_k$ be an ear-decomposition with $\varphi(G)$ even ears.
Apply Lemma~\ref{earinduction}(a) to the ears $P_k,\ldots,P_1$ (in reverse order).
Summing up the obtained inequalities, we get the claim.
\endproof

The number $|V(G)|+\varphi(G)-1$ is even, since an even ear adds an odd number of vertices.
The bound of the Proposition is tight for every 2-edge-connected graph $G$ in the following sense:

\begin{theorem}[\cite{Fra93}]
\label{thm:frank}
Let $G$ be a $2$-edge-connected graph. Then there exists $T\subseteq V(G)$, $|T|$ even, such that
$\tau(G,T)=\frac{1}{2}(|V(G)|+\varphi(G)-1)$.
Such a $T$ and an ear-decomposition with $\varphi(G)$ even ears can be found in $O(|V(G)||E(G)|)$ time.
\end{theorem}

Now we prove a similar statement to Proposition~\ref{prop:frank} for connected-$T$-joins:

\begin{proposition}
\label{prop:connected}
Let $G$ be a $2$-edge-connected graph and an ear-decomposition of $G$
with $\varphi(G)$ even ears, among which there are $\pi_2$ 2-ears.
Then for every $T\subseteq V(G)$, $|T|$ even,
a connected-$T$-join with at most
$$\textstyle
\frac{3}{2}(|V(G)|-1)+ \pi_2 - \frac{1}{2}\varphi(G)$$
edges can be found in $O(|E(G)|)$ time.
\end{proposition}

\prove
Apply  Lemma~\ref{earinduction}(b) to the nontrivial ears in reverse order.
Summing up the obtained inequalities, we get a connected-$T$-join with at most
$\frac{3}{2}(|V(G)|-1)+ \frac{1}{2}\varphi(G) - l$ edges,
where $l$ is the number of nontrivial ears that are not short.
Note that $l$ is at least the number of even ears that are not short,
that is, at least $\varphi(G) - \pi_2$.
The claim follows.
\endproof

\subsection{Nice ear-decompositions}

We need ear-decompositions with particular properties:

\begin{definition}
\label{defnice}
Let $G$ be a graph.
An ear-decomposition of $G$ is called {\em nice} if
\vspace{-5pt}
\begin{itemize}
\addtolength{\itemsep}{-6pt}
\item[{\rm (i)}] the number of even ears is $\varphi(G)$;
\item[{\rm (ii)}] all short ears are pendant;
\item[{\rm (iii)}] internal vertices of different short ears are non-adjacent in $G$.
\end{itemize}
\vspace{-5pt}

\smallskip
An {\em eardrum} in $G$ is the set $M$ of components of an induced subgraph in which every vertex has degree at most $1$.
Let $V_M:=\bigcup M$ be the vertex set of this subgraph.
That is, the one-element sets of  $M$  are isolated vertices in $G[V_M]$ and the two-element sets form an induced matching.

Given a nice ear-decomposition and $T\subseteq V(G)$ with $|T|$ even,
we call an ear $P$ {\em clean} if it is short (and thus pendant) and $\inn(P)\cap T=\emptyset$.
We say that $M$ is the eardrum {\em associated} with the ear-decomposition and $T$
if $M$ is the set of components of the subgraph induced by the set
of internal vertices of the clean ears.
\end{definition}

Another way of saying (iii): the components of the subgraph induced by the internal vertices of short ears form an eardrum
(that is, the only edges in this induced subgraph are the middle edges of  $3$-ears).
The following is essentially Proposition 4.1 of \cite{CheSS01}:

\begin{lemma}
\label{newniceeardecomp}
For any $2$-vertex-connected graph $G$ there exists
a nice ear-decomposition,
and such an ear-decomposition can be computed in $O(|V(G)||E(G)|)$ time.
\end{lemma}

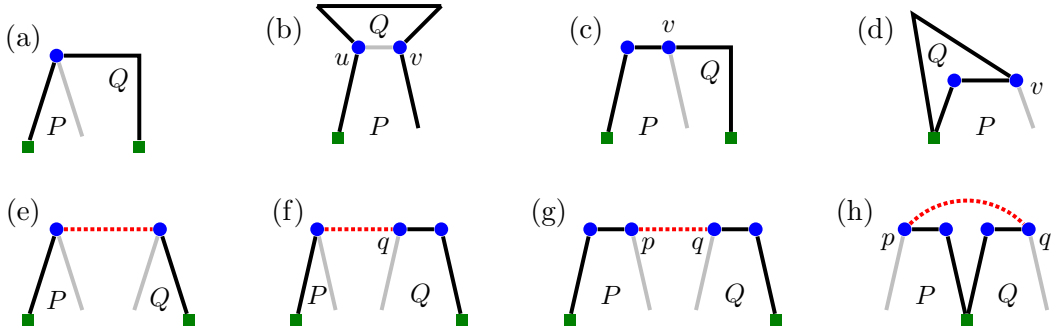
\begin{figure}[ht]
\begin{center}
 \begin{tikzpicture}[scale=0.55]

  \tikzstyle{vertex}=[blue,circle,fill,minimum size=5.2,inner sep=0pt]
  \tikzstyle{endvertex}=[darkgreen,fill,minimum size= 4.5,inner sep=0pt]
  \tikzstyle{edge}=[line width=1.5]
  \tikzstyle{path}=[line width=1.5]
  \tikzstyle{connect}=[red, line width=1.5, densely dotted]
  \tikzstyle{deleteedge}=[grey, line width=1.5]


 \begin{scope}[shift={(0.5,0)}]
 \node[endvertex] (a0)  at ( 0.8, 0) {};
 \node[vertex] (a1)  at ( 1.5, 2.2) {};
 \node (a2)  at ( 2.2, 0) {};
 \node[endvertex] (a4)  at ( 3.5, 0) {};

 \draw[edge] (a0) to (a1);
 \draw[deleteedge] (a1) to (a2);
 \draw[path] (a1) -| (a4);

 \node at (1.5,0.5) {\small $P$};
 \node at (3,1.6) {\small $Q$};
 \node at (0.7,2.6) {(a)};
 \end{scope}

 \begin{scope}[shift={(-5.2,4.4)}]

 \node[endvertex] (b3)  at ( 14, -4.2) {};
 \node[vertex] (b4)  at ( 14.5, -2) {};
 \node[vertex] (b5)  at ( 15.5, -2) {};
 \node (b6)  at ( 16, -4.2) {};

 \draw[edge] (b3) to (b4);
 \draw[deleteedge] (b4) to (b5);
 \draw[edge] (b5) to (b6);
  \draw[edge] (b4) to (13.5,-1);
 \draw[edge] (13.5,-1) to (16.5,-1);
 \draw[edge] (16.5,-1) to (b5);

 \node at (15.0,-3.9) {\small $P$};
 \node at (15.0,-1.5) {\small $Q$};
 \node at (14.1,-2.3) {\small $u$};
 \node at (15.9,-2.3) {\small $v$};
 \node at (12.7,-1.6) {(b)};
 \end{scope}

 \begin{scope}[shift={(-2.7,4.4)}]

 \node[endvertex] (w0)  at ( 18, -4.2) {};
 \node[vertex] (w1)  at ( 18.5, -2) {};
 \node[vertex] (w2)  at ( 19.5, -2) {};
 \node (w3)  at ( 20, -4.2) {};
 \node[endvertex] (w7)  at ( 21.0, -4.2) {};

 \draw[edge] (w0) to (w1);
 \draw[edge] (w1) to (w2);
 \draw[deleteedge] (w2) to (w3);
 \draw[path] (w2) -| (w7);

 \node at (19,-3.9) {\small $P$};
 \node at (20.5,-2.6) {\small $Q$};
 \node at (19.5,-1.5) {\small $v$};
 \node at (17.5,-1.6) {(c)};
 \end{scope}

 \begin{scope}[shift={(5.2,4.4)}]

 \node[endvertex] (w0)  at ( 18, -4.2) {};
 \node[vertex] (w1)  at ( 18.5, -2.8) {};
 \node[vertex] (w2)  at ( 20, -2.8) {};
 \node (w3)  at ( 20.5, -4.2) {};

 \draw[edge] (w0) to (w1);
 \draw[edge] (w1) to (w2);
 \draw[deleteedge] (w2) to (w3);
 \draw[path] (w2) to (17.5,-1.2) to (w0);

 \node at (19.25,-3.9) {\small $P$};
 \node at (18.1,-2.2) {\small $Q$};
 \node at (20.5,-3) {\small $v$};
 \node at (16.6,-1.6) {(d)};
 \end{scope}


 \begin{scope}[shift={(-4.5,-4.2)}]
 \node[endvertex] (b0)  at ( 5.8, 0) {};
 \node[vertex] (b1)  at ( 6.5, 2.2) {};
 \node (b2)  at ( 7.2, 0) {};
 \node (b3)  at ( 8.3, 0) {};
 \node[vertex] (b4)  at ( 9, 2.2) {};
 \node[endvertex] (b5)  at ( 9.7, 0) {};

 \draw[edge] (b0) to (b1);
 \draw[deleteedge] (b1) to (b2);
 \draw[connect] (b1) to (b4);
 \draw[deleteedge] (b3) to (b4);
 \draw[edge] (b4) to (b5);

 \node at (6.5,0.5) {\small $P$};
 \node at (9,0.5) {\small $Q$};
 \node at (5.7,2.6) {(e)};
 \end{scope}

 \begin{scope}[shift={(-9.7,-4.2)}]

 \node[endvertex] (b0)  at ( 17.5, 0) {};
 \node[vertex] (b1)  at ( 18, 2.2) {};
 \node (b2)  at ( 18.5, 0) {};
 \node (b3)  at ( 19.5, 0) {};
 \node[vertex] (b4)  at ( 20, 2.2) {};
 \node[vertex] (b5)  at ( 21, 2.2) {};
 \node[endvertex] (b6)  at ( 21.5, 0) {};

 \draw[edge] (b0) to (b1);
 \draw[connect] (b1) to (b4);
 \draw[deleteedge] (b1) to (b2);
 \draw[deleteedge] (b3) to (b4);
 \draw[edge] (b4) to (b5);
 \draw[edge] (b5) to (b6);

 \node at (19.6,1.8) {\small $q$};
 \node at (18,0.6) {\small $P$};
 \node at (20.5,0.6) {\small $Q$};
 \node at (17.3,2.6) {(f)};
 \end{scope}

 \begin{scope}[shift={(7.4,0)}]

 \node[endvertex] (v0)  at ( 7, -4.2) {};
 \node[vertex] (v1)  at ( 7.5, -2) {};
 \node[vertex] (v2)  at ( 8.5, -2) {};
 \node (v3)  at ( 9, -4.2) {};
 \node (v4)  at ( 10, -4.2) {};
 \node[vertex] (v5)  at ( 10.5, -2) {};
 \node[vertex] (v6)  at ( 11.5, -2) {};
 \node[endvertex] (v7)  at ( 12, -4.2) {};

 \draw[edge] (v0) to (v1);
 \draw[edge] (v1) to (v2);
 \draw[deleteedge] (v2) to (v3);
 \draw[deleteedge] (v4) to (v5);
 \draw[edge] (v5) to (v6);
 \draw[edge] (v6) to (v7);
 \draw[connect] (v2) to (v5);

 \node at (8,-3.6) {\small $P$};
 \node at (11,-3.6) {\small $Q$};
 \node at (8.9,-2.4) {\small $p$};
 \node at (10.1,-2.4) {\small $q$};
 \node at (6.5,-1.6) {(g)};
 \end{scope}

 \begin{scope}[shift={(15,0)}]

 \node (v0)  at ( 7, -4.2) {};
 \node[vertex] (v1)  at ( 7.5, -2) {};
 \node[vertex] (v2)  at ( 8.5, -2) {};
 \node[endvertex] (v3)  at ( 9, -4.2) {};
 \node[vertex] (v5)  at ( 9.5, -2) {};
 \node[vertex] (v6)  at ( 10.5, -2) {};
 \node (v7)  at ( 11, -4.2) {};

 \draw[deleteedge] (v0) to (v1);
 \draw[edge] (v1) to (v2);
 \draw[edge] (v2) to (v3);
 \draw[edge] (v3) to (v5);
 \draw[edge] (v5) to (v6);
 \draw[deleteedge] (v6) to (v7);
 \draw[connect] (v1) to[out=45,in=135] (v6);

 \node at (8,-3.6) {\small $P$};
 \node at (10,-3.6) {\small $Q$};
 \node at (7.1,-2.3) {\small $p$};
 \node at (10.9,-2.3) {\small $q$};
 \node at (6.3,-1.6) {(h)};
 \end{scope}

 \end{tikzpicture}
 \caption{\label{figniceear}
{\small  Proof of Lemma \ref{newniceeardecomp}.
 Squares and circles represent distinct vertices; moreover, vertices represented by
 circles  are internal vertices of short, pendant ears.
 Grey edges become 1-ears. } }
\end{center}
\end{figure}

\prove
Take any open ear-decomposition with $\varphi(G)$ even ears.
This can be done by Proposition~3.2 of \cite{CheSS01}. (Its proof, briefly:
start with Theorem~\ref{thm:frank}, then subdivide an arbitrary edge on each even ear,
apply Theorem 5.5.2 of \cite{LPL} to construct an open odd ear-decomposition
of this $2$-connected factor-critical graph; finally undo the subdivisions.)

We will now satisfy the conditions (ii) and (iii) by successively modifying the ear-decomposition.
Each of the operations that we will use decreases the number of nontrivial ears,
and does not increase the number of even ears.
Moreover pendant ears vanish or remain pendant in each operation.

First we make all 2-ears pendant.
If a $2$-ear $P$ is not pendant,
let $Q$ be the first nontrivial ear attached to it
(Figure \ref{figniceear}(a)).
Then we can replace $P$ and $Q$  by the ear $Q+e$ and the 1-ear $e'$,
where $\{e,e'\}=E(P)$, and $e$ is chosen so that $Q+e$ is open.
The new nontrivial ear $Q+e$ can be put at the place of $Q$ in the ear-decomposition.

Next we make all 3-ears pendant.
As long as this is not the case, we do the following.
Let $P$ be the first non-pendant 3-ear, and let
$Q$ be the first nontrivial (open) ear attached to $P$.
Let $\inn(P)=\{u,v\}$, and let $v$ be an endpoint of $Q$.
If the other endpoint of $Q$ is $u$, then we can form an ear $R$ with
$E(R)=E(Q)\cup E(P) \setminus \{\{u,v\}\}$ (Figure \ref{figniceear}(b)).
Otherwise we form $R$ by $Q$ plus the 2-subpath of $P$ ending in $v$ (Figure \ref{figniceear}(c),(d)).
We replace $P$ and $Q$ by $R$ and a new 1-ear.
The new nontrivial ear $R$ has length at least 4; it can be open or closed.
It can be put at the place of $Q$ in the ear-decomposition.
Since $P$ was the first non-pendant 3-ear, we maintain the property
that no closed ear is attached to any 3-ear.

\medskip

Now all short ears are pendant.
This also implies that there are no edges connecting internal vertices of 2-ears:
otherwise one could replace the two (pendant) 2-ears and the 1-ear connecting them
by an open pendant 3-ear and two 1-ears (Figure \ref{figniceear}(e)),
reducing the number of even ears by two.

We still have to obtain property (iii).
If there is an edge $e$ that connects the internal vertex of a 2-ear $P$
with an internal vertex $q$ of a 3-ear $Q$,
let $Q'$ be the 2-subpath of $Q$ with endpoint $q$.
Form a new open 4-ear $R$ by $Q'$, $e$, and one edge of $P$ (Figure \ref{figniceear}(f)).
We replace $P$, $Q$, and the 1-ear consisting of $e$ by $R$ and two new 1-ears.
The new nontrivial ear $R$ is pendant, so it can be put at the end of the ear-decomposition,
followed only by 1-ears.

Finally, if there is an edge $e=\{p,q\}$ that connects internal vertices of two different 3-ears $P$ and $Q$,
we form a new 5-ear $R$ by the edge $e$
and the 2-subpaths of $P$ and $Q$ ending in $p$ and $q$ respectively (Figure \ref{figniceear}(g),(h)).
We replace $P$, $Q$, and the 1-ear consisting of $e$
by $R$ and two new 1-ears.
Note that $R$ can be open or closed, but it is always pendant,
so it can be put at the end of the ear-decomposition, followed only by 1-ears.

Since the number of nontrivial ears decreases by each of these operations,
the algorithm will terminate after less than $|V(G)|$ iterations.
At the end, the ear-decomposition is nice.
\endproof

\subsection{How to switch to nicer ears?}\label{sub:nicer}

Our approximation algorithms will begin by computing a nice ear-decomposition.
Lemma \ref{earinduction}(b) indicates that clean ears
are more expensive than others.
We will make up for this by ``optimizing'' them,
in order to serve best for connectivity.

Consider a graph $G$ with a nice ear-decomposition, and let $M$ be the eardrum associated with it
and the given set $T\subseteq V(G)$.
So $M$ contains a 1-element set $\{v\}$ for each  clean 2-ear, where $v$ is the internal vertex of the $2$-ear,
and a 2-element set $\{v,w\}$ for each clean 3-ear where $\{v,w\}$ is the set of internal vertices of the $3$-ear.
Let again $V_M=\bigcup M$.
Note that $V_M\cap T=\emptyset$.
There may be $1$-ears connecting $V_M$ and $V(G)\setminus V_M$,
and these can be used to replace some of the clean ears  by ``more useful'' clean ears of the same length.

\begin{proposition}\label{prop:reroot}
Let $G$ be a 2-edge-connected graph, and
$T\subseteq V(G)$ with $|T|$ even.
Let a nice ear-decomposition be given,
and let $M$ be the eardrum associated with it and $T$.
For $f\in M$ let $P_f$ be the ear with $f$ as set of internal vertices, and
let $Q_f$ be any path in $G$ in which $f$ is the set of internal vertices.
Then replacing the ears $(P_f)_{f\in M}$ by the ears $(Q_f)_{f\in M}$ and changing the set of 1-ears accordingly,
we get a nice ear-decomposition again with the same associated eardrum.
\end{proposition}

\prove  Since all 2-ears and 3-ears were already pendant,
no new pendant short ears,
except of course the ears $Q_f$ that replace $P_f$  $(f\in M)$,
can arise by this change.
Moreover, no vertex of $V_M$ can be an endpoint of any path $Q_f$ ($f\in M$).
Hence the new ear-decomposition is also nice, and the
eardrum associated with the ear-decomposition and $T$ remains the same.
\endproof

We will choose the paths $Q_f$ $(f\in M)$ such that
$(V(G),\bigcup_{f\in M} E(Q_f) )$ has as few components as possible.
We will show how in the next section.
Let us denote this minimum by $c(G,M)$.
Then adding $c(G,M)-1$ edges to the $|M|+|V_M|$ edges of $\bigcup_{f\in M} E(Q_f)$
yields a connected spanning subgraph in which all vertices in $V_M$ have even degree.
It is not difficult to see (and we will show it in Corollary \ref{cor:connectandeven} below)
that there is no such subgraph with fewer edges.

\section{Earmuffs}\label{sec:earmuff}

Let $G$ be a graph and $M$ an eardrum in $G$.
For each $f\in M$, let $\Pscr_f$ be the set of $(|f|+1)$-paths in $G$ in which $f$ is the set of internal vertices.
In other words, for $|f|=2$ (or $|f|=1$), $\Pscr_f$ is the set of possible 3-ears (or $2$-ears)  containing $f$ as middle edge
(or the unique element of $f$ as middle vertex, respectively).
As explained in Subsection \ref{sub:nicer},
we want to pick an element $P_f\in \Pscr_f$ for each $f\in M$ such that we need to add
as few further edges as possible to the graph $(V(G),\bigcup_{f\in M}E(P_f))$ in order to make it connected.
Ideally, if this graph is a forest, then
$|V(G)|-1-|M| - |V_M|$
further edges suffice.
This motivates the following definitions:

\begin{definition}
Let $G$ be a graph and $M$ an eardrum in $G$.
For $f\in M$ let $\Pscr_f$ denote the set of paths $P$ in $G$ with $\inn(P)=f$.
An {\em earmuff} (for $M$ in $G$) is a set of  paths $\{P_f: f\in F\}$, where $F\subseteq M$ and $P_f\in\Pscr_f$,
such that  $(V(G),\bigcup_{f\in F}E(P_f))$ is a forest.
\end{definition}

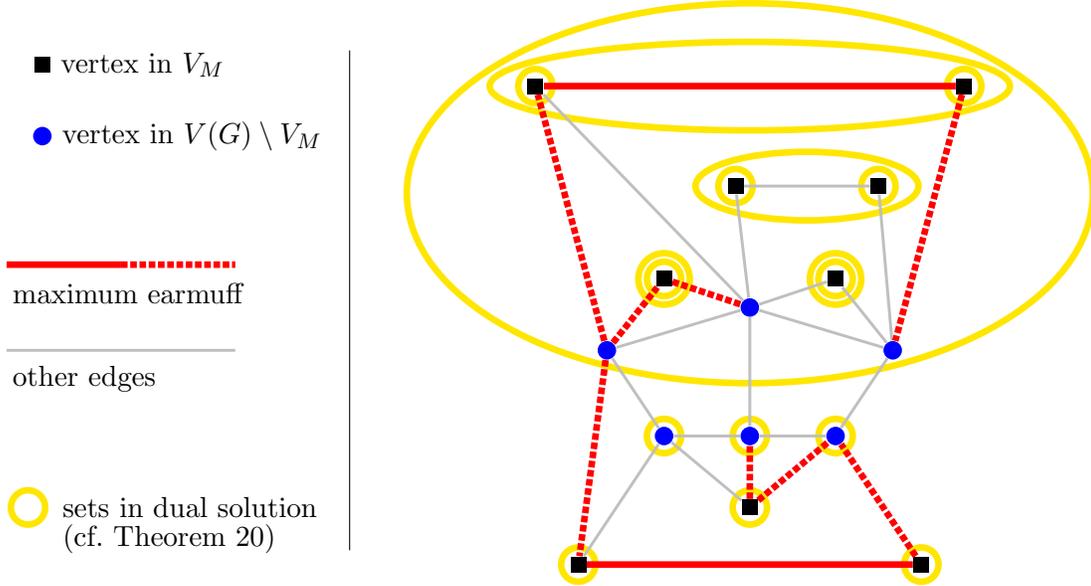
\begin{figure}[t]
\begin{center}
\begin{tikzpicture}[scale=0.19]

 \tikzstyle{vertex}=[blue,circle,fill,minimum size=7,inner sep=0pt]
 \tikzstyle{earvertex}=[black,fill,minimum size=6,inner sep=0pt]
 \tikzstyle{edge}=[grey, line width=1.1, rounded corners]
 \tikzstyle{ear}=[red, line width=2.5, densely dotted, rounded corners]
 \tikzstyle{matching}=[red, line width=2.5, rounded corners]
 \tikzstyle{badmatching}=[grey,line width=1.1, rounded corners]
 \tikzstyle{ignore}=[grey, line width=1.1, rounded corners]
 \tikzstyle{dual}=[orange, line width=2.5]

\node[earvertex, label=right:vertex in $V_M$] at (-19.5,30) {};
\node[vertex, label=right:vertex in $V(G)\setminus V_M$] at (-19.5,25) {};

 \begin{scope}[shift={(-2,0)}]
\draw[matching] (-20,16)--(-12,16);
\draw[ear] (-12,16)--(-4,16);
\node[label=right:maximum earmuff] at (-21,14) {};
\draw[edge] (-20,10)--(-4,10);

\node[label=right:other edges] at (-21,8) {};
\draw[dual] (-18.5,-1) circle (1.2);
\node[label=right:sets in dual solution] at (-17.5,-1) {};
\node[label=right:(cf.\ Theorem~\ref{lbmuT})] at (-17.5,-3.2) {};
\end{scope}
\draw (2,-4)--(2,31);

\node[vertex] (v1)  at ( 20, 10) {};
\node[vertex] (v2)  at ( 30,13) {};
\node[vertex] (v3)  at ( 40,10) {};
\node[earvertex] (v4)  at ( 24,15) {};
\node[earvertex] (v6)  at ( 36,15) {};
\node[earvertex] (v8)  at ( 29,21.5) {};
\node[earvertex] (v9)  at ( 39,21.5) {};
\node[earvertex] (v10)  at ( 15,28.5) {};
\node[earvertex] (v11) at ( 45,28.5) {};
\node[vertex] (v14)  at (24,4) {};
\node[vertex] (v16)  at (36,4) {};
\node[vertex] (v17)  at (30,4) {};
\node[earvertex] (v18)  at (18,-5) {};
\node[earvertex] (v19)  at (42,-5) {};
\node[earvertex] (v20)  at (30,-1) {};

\draw[dual] (v4) circle (1.2);
\draw[dual] (v4) circle (1.8);
\draw[dual] (v6) circle (1.2);
\draw[dual] (v6) circle (1.8);
\draw[dual] (v8) circle (1.2);
\draw[dual] (v9) circle (1.2);
\draw[dual] (v10) circle (1.2);
\draw[dual] (v11) circle (1.2);
\draw[dual] (v14) circle (1.2);
\draw[dual] (v16) circle (1.2);
\draw[dual] (v17) circle (1.2);
\draw[dual] (v18) circle (1.2);
\draw[dual] (v19) circle (1.2);
\draw[dual] (v20) circle (1.2);
\draw[dual] (30,21.0) circle (24 and 13.3);
\draw[dual] (30,28.5) circle (18.2 and 3.1);
\draw[dual] (34,21.5) circle (7.8 and 2.4);

 \draw[edge] (v1) to (v2);
 \draw[edge] (v2) to (v3);
 \draw[ear] (v1) to (v4);
 \draw[ear] (v4) to (v2);
 \draw[edge] (v2) to (v6);
 \draw[edge] (v6) to (v3);
 \draw[edge] (v2) to (v8);
 \draw[badmatching] (v8) to (v9);
 \draw[edge] (v9) to (v3);
 \draw[ear] (v1) to (v10);
 \draw[ignore] (v10) to (v2);
 \draw[matching] (v10) to (v11);
 \draw[ear] (v11) to (v3);
 \draw[edge] (v1) to (v14);
 \draw[edge] (v14) to (v17);
 \draw[edge] (v2) to (v17);
 \draw[edge] (v16) to (v3);
 \draw[edge] (v16) to (v17);
 \draw[ear] (v1) to (v18);
 \draw[ignore] (v14) to (v18);
 \draw[matching] (v18) to (v19);
 \draw[ear] (v19) to (v16);
 \draw[ignore] (v20) to (v14);
 \draw[ear] (v20) to (v17);
 \draw[ear] (v20) to (v16);

\end{tikzpicture}
\caption{\label{figearmuff}
{\small an eardrum, a maximum earmuff, and an optimum dual solution} }
\end{center}
\end{figure}

A {\em maximum earmuff} is one in which $|F|$, its {\em size}, is maximum, and
this maximum is denoted by $\mu(G,M)$.
See Figure \ref{figearmuff} for an illustration.
We show now that a maximum earmuff can be computed in polynomial time.
There are two ways at hand: one uses matroid intersection,
the other one forest representative systems (generalizing bipartite matching).
The first one has a  shorter proof, the second is more elementary,
leads to a faster algorithm,
and may be easier to have in mind for illustrating a dual solution of the LP relaxation.

\subsection{Maximum Earmuffs by Matroid Intersection}

We use the following well-known theorem:

\begin{theorem}[\cite{Rad42}]
\label{rado}
Let $E$ be a finite set and $r$ the rank function of a matroid on $E$.
Let $E_1,E_2,\ldots,E_k\subseteq E$.
Then
$$\max \bigl\{ r(\{e_1,\ldots,e_k\}) : e_i\in E_i \ (i=1,\ldots,k) \bigr\}
\ = \ \textstyle
\min\left\{r \left( \bigcup_{i\in I} E_i \right) + k - |I| : I\subseteq\{1,\ldots,k\} \right\}.
$$
\end{theorem}

It is an easy and well-known exercise to deduce this from the matroid intersection theorem
(\cite{Edm70}).
Therefore
one can find a set attaining the maximum in polynomial time
using the matroid intersection algorithm.

In order to apply Rado's Theorem directly, we represent each path $P\in\Pscr_f$ ($f\in M$) by the set
$e_P\in {V(G)\setminus V_M \choose 2}$ of its two endpoints.
Let $r$ be the rank function of the cycle matroid of the complete graph on $V(G)\setminus V_M$.
If we write $E_f:=\{e_P: P\in\Pscr_f\}$ for $f\in M$, then
$$\mu(G,M) \ = \ \max\{r(\{e_f:f\in M\}) : e_f\in E_f \ (f\in M)\}.$$
Hence we can find a maximum earmuff in polynomial time.

\subsection{Maximum Earmuffs and Forest Representatives}

This section provides an alternative (more elementary and faster) solution
to the earmuff maximization problem.

Let $U$ and $M$ be  finite sets, and let $U_f\subseteq U$ for $f\in M$.
Then $(e_f)_{f\in M}$ is called a {\em forest representative system} for $(U_f)_{f\in M}$ if
$e_f\in {U_f \choose 2}$ for all $f\in M$, $e_f\not=e_{f'}$ for $f\not=f'$,
and the graph $(U,\{e_f:f\in M\})$ is a forest.

\begin{corollary}[\cite{L70}]
\label{frs}
Let $U$ and $M$ be  finite sets, and let $\emptyset\not=U_f\subseteq U$ for $f\in M$.
Then the maximum cardinality of a subset $F\subseteq M$ for which
$(U_f)_{f\in F}$ has a forest representative system
equals
$$\min \biggl\{ |M| - \sum_{W\in\Wscr} \bigl( |\{f\in M: U_f\subseteq W\}|-(|W|-1) \bigr) :
\Wscr \mbox{ is a partition of } U \biggr\}.$$
\end{corollary}

This is a  variant of Corollary 1.4.6 of \cite{LPL},
where bipartite matchings are used in the proof, convertible to an algorithm.
It also follows directly from Rado's Theorem:

\smallskip
\prove
The inequality ``$\le$'' follows from the fact
that for every partition  $\Wscr$ of $U$ and each $W\in\Wscr$
at most $|W|-1$ of the $f\in M$ with $U_f \subseteq W$ can be represented,
and the sets $\{f\in M:U_f\subseteq W\}$ are pairwise disjoint for
different sets $W\in\Wscr$ because all $U_f$ are nonempty.

For the other direction, apply Theorem \ref{rado} to the sets ${U_f \choose 2}$ ($f\in M$)
and the cycle matroid of the complete graph on $U$.
We get a forest representative system of size
$r \bigl( \bigcup_{f\in F} {U_f \choose 2} \bigr)  +|M| - |F|$ for some $F\subseteq M$.
Let $\Wscr$ be the set of components of
the graph $\bigl(U,\bigcup_{f\in F}{U_f \choose 2} \bigr)$.
We have
$r \bigl( \bigcup_{f\in F} {\textstyle{U_f \choose 2}} \bigr)  =  \sum_{W\in\Wscr} (|W|-1)$
and
$|F| \le \sum_{W\in\Wscr} |\{f\in M: U_f\subseteq W\}|$
because, by the definition of $\Wscr$, for every $f\in F$ there is a $W\in\Wscr$ with $U_f\subseteq W$.
\endproof

\medskip

We give now an elementary and algorithmic proof of the nontrivial inequality of Corollary \ref{frs},
giving rise to an efficient algorithm for computing a maximum earmuff in $O(|V(G)||E(G)|)$ time.

Let $F\subseteq M$ such that $(U_f)_{f\in F}$ has a forest representative system $(e_f)_{f\in F}$.
A set $W\subseteq U$ will be called {\em $F$-closed} if $|\{f\in F: U_f\subseteq W\}|=|W|-1$.
For any $F$-closed set $W$, the graph $(W,\{e_f:f\in F,\ U_f\subseteq W\})$ is a tree.
Therefore the union of two $F$-closed sets with nonempty intersection is also $F$-closed.
Moreover, every singleton is $F$-closed.
We conclude that the set of maximal $F$-closed sets is a partition of $U$.

If $F$ is a maximum subset of $M$ such that $(U_f)_{f\in F}$ has a forest representative system,
then this partition certifies maximality, as we shall prove now.

\begin{lemma}
\label{frsalgorithmlemma}
Let $U$ and $M$ be finite sets, and let $U_f\subseteq U$ for $f\in M$.
Let $F\subseteq M$ and a forest representative system $(e_f)_{f\in F}$
for $(U_f)_{f\in F}$ be given, and let $g\in M\setminus F$.
Then one can
\vspace{-5pt}
\begin{itemize}
\addtolength{\itemsep}{-6pt}
\item[--] either
find
a forest representative system $(e'_f)_{f\in F\cup \{g\}}$
for $(U_f)_{f\in F\cup\{g\}}$
\item[--] or conclude that $U_g$ is contained in an $F$-closed set
\end{itemize}
\vspace{-5pt}
in  $O(\sum_{f\in M}|U_f|)$ time.
\end{lemma}

\medskip
\prove
Let $F\subseteq M$ and a forest representative system $(e_f)_{f\in F}$ for $(U_f)_{f\in F}$ be given.
Let $E_F:=\{e_f: f\in F\}$, and consider the forest $(U,E_F)$.
Let $\Cscr$ be the set of components of $(U,E_F)$.
Let $T:=\{f\in M : U_f\not\subseteq C \mbox{ for all } C\in\Cscr\}$.
Consider the digraph $D$ on the vertex set $M$ that contains an edge
$(f,f')$ if and only if $f\in M\setminus T$, $f'\in F$, and there exist $u,v\in U_f$ such that $e_{f'}$ lies
on the unique $u$-$v$-path in $(U,E_F)$.
We call $f$ {\em reachable from} $g$ if  there exists a directed path $P$ from $g$ to $f$ in $D$.

\medskip
\boldheader{Claim~1}
If there is an $f\in T$ that is reachable from $g$,
then $F\cup\{g\}$ has a forest representative system.

\smallskip
To prove this, let $P$ be a shortest directed path from $g$ to $f\in T$ in $D$.
Let $g=f_0,f_1,\ldots,f_k=f$ be the vertices of $P$ in this order.
Set $e'_f:=e_f$ for all $f\in F\setminus\{f_0,\ldots,f_k\}$.

Let $e'_f$ be a pair $\{u_f,v_f\}$ such that $v_f$ is not in the same component of $(U,E_F)$ as $u_f$.
For each arc $a=(f_i,f_{i+1})$ of $P$ we have $u_i,v_i\in U_{f_i}$ such that $e_{f_{i+1}}$
(but no $e_{f_j}$ with $j>i+1$)
lies on the unique $u_i$-$v_i$-path in $(U,E_F)$, and we set $e'_{f_i}:=\{u_i,v_i\}$.
A straightforward induction shows that
$(U,\{e_f:f\in F\setminus\{f_{j+1},\ldots,f_k\}\}\cup\{e'_{f_j},\ldots,e'_{f_k}\})$
is a forest for all $j=k,k-1,\ldots,0$.
For $j=0$ this means that
$(U,\{e'_f:f\in F\cup\{g\})$ is a forest, and
Claim~1 is proved.

\medskip
\boldheader{Claim~2}
If no element of $T$ is reachable from $g$, then
$U_g$ is contained in an $F$-closed set.

\smallskip
Indeed,
if $R$ is the set of vertices that
are reachable from $g$ in $D$, and $R\cap T=\emptyset$,
then $\bigcup \{U_f: f\in R\}$ is $F$-closed.

\medskip
The two Claims directly imply an algorithm:
we perform a BFS search from $g$ in $D$.
To do this efficiently, we  fix an element $r\in U_g$ (we may assume that $U_g$ is nonempty),
compute the components of $(U,E_F)$,
and orient the component $C$ containing $r$ as an arborescence rooted at $r$.
We work with a queue $Q$ that we initialize so that it contains only $g$,
and do the following until we reach an element of $T$ or cannot continue because $Q$ is empty.

Remove the first element $f$ from $Q$.
For all $u\in U_f$, check whether $u\in C$ (if not, $f\in T$, and we are done)
and traverse the $u$-$r$-path in $(U,E_F)$
(always following the incoming arc in the arborescence)
as long as we visit edges that we have not visited before.
For each such edge $e_{f'}$ we insert $f'$ at the end of the queue $Q$
and store that $f$ was the predecessor of $f'$.

Note that the set of visited edges always forms a tree containing $r$.
If $f'$ enters the queue with predecessor $f$, then $(f,f')$ is an arc of $D$.
The correctness and the claimed running time follow.
 \endproof

\begin{theorem}
\label{frsalgorithm}
Let $U$ and $M$ be finite sets, and let $U_f\subseteq U$ for $f\in M$.
A maximum subset $F\subseteq M$ with a forest representative system for $(U_f)_{f\in F}$
can be computed in
$O(|M|\sum_{f\in M}|U_f|)$ time.
\end{theorem}

\prove
We may assume $U_f\not=\emptyset$ for all $f\in M$.
Let $M=\{g_1,\ldots,g_n\}$.
We run the greedy algorithm, beginning with $F_0=\emptyset$.
For $j=1,\ldots,n$ we apply Lemma \ref{frsalgorithmlemma}
to $F_{j-1}$, $M_j:=\{g_1,\ldots,g_j\}$, and $g_j$.
We either augment $F_j:=F_{j-1}\cup\{g_j\}$, or we set $F_j:=F_{j-1}$.
In each case we have a forest representative system of $(U_f)_{f\in F_j}$ and
the property that
$U_f$ is contained in an $F_j$-closed set for all $f\in M_j \setminus F_j$.
So each $U_f$ ($f\in M_j\setminus F_j$) is also contained in an element of $\Wscr$,
where $\Wscr$ is the set of maximal $F_j$-closed sets,
and we have
$$|M_j\setminus F_j| \ = \ \sum_{W\in\Wscr} |\{f\in M_j\setminus F_j: U_f\subseteq W\}|.$$
Since all elements of $\Wscr$ are $F_j$-closed, this implies
$$|M_j\setminus F_j| \ = \ \sum_{W\in\Wscr} \bigl( |\{f\in M_j: U_f\subseteq W\}| - (|W|-1)\bigr).$$
By the trivial inequality of Corollary \ref{frs},
this implies that $F_j$ is a maximum subset of $M_j$ with a forest representative system.
\endproof

This is an algorithmic reformulation of
the following result of \cite{Lor75} (see \cite{Fra11} for a direct proof):
given a hypergraph, the sets of hyperedges that have a forest representative system form the independent sets of a matroid.

\bigskip
We now apply forest representative systems to compute a maximum earmuff.

\bigskip
Let $M$ be an eardrum in $G$, and let $U:=V(G)\setminus V_M\not=\emptyset$. We will denote by
$U_f$  the set of endpoints of paths in $\Pscr_f$ ($f\in M$).
For $W\subseteq V(G)\setminus V_M$ we define the {\em surplus} of $W$ as
$\deff(W):= |\{f\in M: U_f\subseteq W\}|-(|W|-1)$.
In particular, if $|W|=1$, then $\deff(W)=0$.

\begin{lemma}
\label{frsearmuff}
$\mu(G,M)$ is the maximum cardinality of a subset $F\subseteq M$ for which
$(U_f)_{f\in F}$ has a forest representative system.
Given a forest representative system, we can compute an earmuff of the same
size in $O(|V(G)|^2)$ time.
\end{lemma}

\prove
Given an earmuff with $F\subseteq M$ and $P_f\in\Pscr_f$ for $f\in F$,
then $\{e_{P_f}:f\in F\}$ is a forest representative system for $(U_f)_{f\in F}$.

Conversely, let $\{e_f:f\in F\}$ be a forest representative system for $(U_f)_{f\in F}$.
We will successively replace each $e_f$ ($f\in F$) by the edge set of a path
$P_f\in\Pscr_f$ and maintain a forest.

So let $f\in M$.
Since $e_f\in {U_f \choose 2}$, say $e_f=\{u,v\}$, there are paths $P,Q\in\Pscr_f$ such that
$u$ is an endpoint of $P$ and $v$ is an endpoint of $Q$.

If $|f|=1$, say $f=\{a\}$, then $a$ is adjacent to $u$ (in $P$, and thus in $G$) and to $v$ (in $Q$, and thus in $G$).
So let $P_f$ be the 2-path with vertices $u,a,v$ in this order.

If $|f|=2$, suppose that the vertices of $P$ are $u,a,b,w$ in this order.
Note that $v$ is adjacent to $a$ or $b$ (in $Q$, and thus in $G$).

If $v$ is adjacent to $b$, then
let $P_f$ be the 3-path with vertices $u,a,b,v$ in this order.
If $v$ is adjacent to $a$, then consider the path $R$ with vertices $v,a,b,w$ in this order.
Since the edge $e_f$ (as every edge in a forest) is a bridge,
we can choose $P_f$ as one of $P$ or $R$ and replace $e_f$ by $E(P_f)$ without creating a circuit.
\endproof

We conclude:

\begin{theorem}
\label{earmuffmain}
Let $G$ be a graph and $M$ an eardrum in $G$ with $\Pscr_f\not=\emptyset$ for all $f\in M$.
Then a maximum earmuff can be computed in $O(|V(G)||E(G)|)$ time, and its size is
$$\mu(G,M) \ = \ \min \biggl\{ |M| - \sum_{W\in\Wscr} \deff(W) :
\Wscr \mbox{ is a partition of } V(G)\setminus V_M \biggr\}.$$
\end{theorem}

\prove
Follows directly from Corollary~\ref{frs}, Theorem~\ref{frsalgorithm}, and Lemma~\ref{frsearmuff}.
\endproof

\section{Lower Bounds \label{sclb}}

To prove the approximation guarantees of our algorithms, we need several lower bounds.

\begin{theorem}[\cite{CheSS01}]
\label{lbphi}
Let $G$ be a 2-edge-connected graph. Then
$$L_{\varphi}(G) \ := \ |V(G)|+\varphi(G)-1 \ \le \ \lp(G).$$
In particular, every 2-edge-connected spanning subgraph of $G$ has at least
$L_{\varphi}(G)$ edges.
\end{theorem}

\prove
By Theorem \ref{thm:frank} there exists a $T\subseteq V(G)$ with $|T|$ even
such that $\frac{1}{2}L_{\varphi}(G)$ is the minimum cardinality of a $T$-join in $G$.
By a well-known result due to \cite{EdmJ73} and \cite{L75},
this implies that there exists a multiset of $L_{\varphi}(G)$ $T$-cuts containing every edge at most twice.
By summing the inequalities $x(\delta(W))\ge 2$ for all these cuts, we obtain
$\lp(G) \ge  L_{\varphi}(G)$.
\endproof

Consequently $L_{\varphi}(G)\le\ec(G)$, and this can indeed be seen more easily: it holds
since the number of even ears is at most the number of nontrivial ears in any ear-decomposition.

Recall that $\lp(G)$ is not a valid lower bound for the connected-$T$-join problem,
and nor are $L_{\varphi}(G)$ and $|V(G)|$.
We use Proposition~\ref{proplpT} and our ``earmuff theorem'' (Theorem~\ref{earmuffmain}) to establish another lower bound:

\begin{theorem}
\label{lbmuT}
Let $G$ be a connected graph, $T\subseteq V(G)$ with $|T|$ even, and $M$ an eardrum in $G$ with $V_M\cap T=\emptyset$
and $\Pscr_f\not=\emptyset$ for all $f\in M$.
Then
$$L_{\mu}(G,M) \ := \ |V(G)|-1+|M|-\mu(G,M) \ \le \ \lp(G,T).$$
In particular, every connected-$T$-join of $G$ has at least $L_{\mu}(G,M)$ edges.
\end{theorem}

\prove
We use Theorem \ref{earmuffmain}.
Let $\Wscr$ be a partition of $V(G)\setminus V_M$ such that
$$\mu(G,M) \ = \
|M| - \sum_{W\in\Wscr} \deff(W).$$

Let $I$ be the subset of $M$ containing those sets $f$ for which
$U_f\subseteq W$ for some $W\in\Wscr$.
Consider the partition $\hat\Wscr$ of $V(G)$ that contains
\vspace{-7pt}
\begin{itemize}
\addtolength{\itemsep}{-6pt}
\item[-] the set $W\cup \bigcup_{f\in M: U_f\subseteq W} f$ for each $W\in\Wscr$;
\item[-] the set $\{x\}$ for each $x\in f \in M\setminus I$.
\end{itemize}
\vspace{-6pt}

\noindent
Next, consider the following multiset $\Sscr$ of nonempty proper subsets of $V(G)$:
\vspace{-7pt}
\begin{itemize}
\addtolength{\itemsep}{-6pt}
\item[-] for each $x\in f\in I$, take the set $\{x\}$;
\item[-] for each $f\in I$, take the set $f$.
\end{itemize}
\vspace{-6pt}

 \noindent
 See Figure \ref{figearmuff} for an illustration.
 Note that singletons in $I$ appear and are counted twice in $\Sscr$.
Each of the sets of $\Sscr$ induces a cut.
 None of these cuts contains an edge of $\delta(\hat\Wscr)$.
 Moreover, no edge belongs to more than two of these cuts.

 Therefore every feasible solution $x$ of $\lp(G,T)$ satisfies
 \begin{eqnarray*}
 x(E(G))
 &=& x(\delta(\hat\Wscr)) \ + \ x(E(G)\setminus\delta(\hat\Wscr)) \\
 &\ge& x(\delta(\hat\Wscr)) \ + \ \frac{1}{2}\sum_{S\in\Sscr} x(\delta(S)) \\
 &\ge& |\hat\Wscr|-1 \ + \ |\Sscr| \\
 &=& |\Wscr|-1 \ + \ |V_M| \ + \ |I| \\
 &=& |\Wscr|-1 \ + \ |V_M| \ + \ \sum_{W\in\Wscr} (\deff(W) + |W|-1) \\
 &=&  |V(G)|-1 + \sum_{W\in\Wscr} \deff(W) \\
 &=& L_{\mu}(G,M).
 \end{eqnarray*}
\mathendproof

For the special case $T=\emptyset$ we note:

\begin{corollary}
\label{lbmu}
Let $G$ be a 2-edge-connected graph and $M$ an eardrum in $G$ with $\Pscr_f\not=\emptyset$ for all $f\in M$.
Then
$$L_{\mu}(G,M) \ \le \ \lp(G).$$
In particular, every 2-edge-connected spanning subgraph of $G$ has at least $L_{\mu}(G,M)$ edges.
\end{corollary}

\prove
This follows from Theorem \ref{lbmuT} and $\lp(G,\emptyset)=\lp(G)$.
\endproof

\smallskip
The following statement will not be explicitly used but may be worth mentioning:

\begin{corollary}
\label{cor:connectandeven}
Let $G$ be a 2-edge-connected graph, and $T\subseteq V(G)$ with $|T|$ even.
Let a nice ear-decomposition be given, and let $M$ be the eardrum associated with it and $T$.
Then $L_\mu(G,M)$ is the minimum number of edges of a connected spanning subgraph of $2G$
in which every vertex of $V_M$ has even degree.
\end{corollary}

\prove
Let $(P_f)_{f\in F}$ be a maximum earmuff for $M$ in $G$, and
for $f\in M\setminus F$ let $P_f$ be the ear with internal vertices $f$.
Taking all the $|M|+|V_M|$ edges in $\bigcup_{f\in M}E(P_f)$
results in a subgraph of $G$ with $|V(G)|-|V_M|-|F|$ components, and every vertex of $V_M$ has even degree.
Adding $|V(G)|-|V_M|-|F|-1$ edges of $G-V_M$ makes the graph connected.
We have used $|M|+|V_M|+|V(G)|-|V_M|-|F|-1=L_\mu(G,M)$ edges in total.

For the converse, Proposition~\ref{proplpT} and
Theorem~\ref{lbmuT} establish $\opt(G,T)\ge \lp(G,T)\ge L_\mu (G,M)$
for all $T\subseteq V(G)$ with $T\cap V_M=\emptyset$.
Thus also the minimum is at least $L_\mu(G,M)$.
\endproof

\smallskip
We will repeat this construction in a similar way
in the first part of the proof of Theorem~\ref{connTjoinbyearmuff}.

\section{Approximation Algorithms \label{scalg}}

All our approximation algorithms begin by computing a suitable ear-decomposition:

\begin{lemma}\label{lem:new}
Let $G$ be a $2$-vertex-connected graph, and $T\subseteq V(G)$ with $|T|$ even.
Then $G$ has a nice ear-decomposition
containing a maximum earmuff
for the eardrum associated with it and $T$.
Such an ear-decomposition can be computed in $O(|V(G)||E(G)|)$ time.
\end{lemma}

\prove
Lemma \ref{newniceeardecomp} provides us with a nice ear-decomposition.
Let $M$ be the eardrum associated with this ear-decomposition and $T$.
Compute a maximum earmuff $(Q_f)_{f\in F}$ $(F\subseteq M)$
for $M$ in $G$ (cf.\ Theorem \ref{earmuffmain}).
Let $(P_f)_{f\in F}$ be the original ears containing the elements of $F$.
Change now the current ear-decomposition by replacing the ears
$(P_f)_{f\in F}$ by  $(Q_f)_{f\in F}$.
By Proposition~\ref{prop:reroot}, the new ear-decomposition is nice, and
the associated eardrum remains the same.
Moreover, the new ear-decomposition contains a maximum earmuff for $M$.
\endproof

\subsection{3/2-approximation for connected-{\boldmath $T$}-joins}\label{sec:connT}

Before describing our three approximation algorithms,
we first prove a theorem for connected-$T$-joins that will be applied for all the three problems
in the case when there are many pendant ears.
``Many" is not the same quantity  though for the three problems.

We have the important inequality $L_\mu(G,M) \le\lp(G,T)\le \opt(G,T)$, for all $T$.
For $T=\emptyset$ this provides a lower bound for $\opt(G)$ and $\ec(G)$ as well.
$L_\varphi(G)$
is also a lower bound for $\ec(G)$ and consequently for $\opt(G)$,
but not for $\opt(G,T)$ in general.
Nevertheless the following
can then also be used in another way.

\begin{theorem}
\label{connTjoinbyearmuff}
Let $G$ be a graph
and $T\subseteq V(G)$ with $|T|$ even,
given with a nice ear-decomposition of $G$
containing a maximum earmuff for
the eardrum $M$ associated with it and $T$.
Then a connected-$T$-join of cardinality at most
$\textstyle
L_\mu(G,M)  +    \frac{1}{2}L_\varphi(G)   -  \pi$
can be constructed in $O(|V(G)|^3)$ time,
where $\pi$ is the number of pendant ears.
\end{theorem}

\prove
Let  $V_M=\bigcup M$ be the set of internal vertices of clean ears.
Define $V_1$ to be the set of internal vertices of pendant but not clean ears,
and $V_0=V(G)\setminus(V_1\cup V_M)$.
Note that $G[V_0]$ is 2-edge-connected.
Let $\varphi_M$ be the number of clean $2$-ears,
$\varphi_1$ the number of even pendant ears that are not clean,
and $\varphi_0=\varphi(G[V_0])$  the number of remaining even ears.
Note that
$\varphi(G)=\varphi_0+\varphi_1+\varphi_M$.

First, let $E_1$ denote the union of the edge sets of the clean ears.
Since these contain a maximum earmuff,
 $(V_M\cup V_0,E_1)$ has $|V_0|-\mu(G,M)$ components.
Note that $|E_1|=\frac{3}{2}|V_M|+\frac{1}{2}\varphi_M$.

Second, we add a set $E_2$ of $|V_0|-\mu(G,M)-1$ edges of $G[V_0]$ such that
$(V_M\cup V_0,E_1\cup E_2)$ is connected.

Third, we apply Lemma \ref{earinduction}(b) to all the remaining $\pi-|M|$ pendant ears.
For each such ear $P$ we add the corresponding edge set $F'$.
Let $E_3$ denote the union of these sets.
Now by Lemma \ref{earinduction},
$(V(G),E_1\cup E_2\cup E_3)$ is connected,
and for each such ear $P$ we added at most $\frac{3}{2}|\inn(P)|+\frac{1}{2}\varphi(P)-1$ edges
(since $\gamma(P)=0$),
so in total
$|E_3| \ \le \
\textstyle
\frac{3}{2}|V_1|+ \frac{1}{2} \varphi_1 - (\pi - |M|).$

Finally, we have to correct the parities of the vertices in $V_0$.
Let $T_0$ be the set of vertices $v\in V_0$ for which $|(E_1\cup E_2\cup E_3)\cap\delta(v)|$
does not have the correct parity (odd if $v\in T$ and even if $v\notin T$).
We add a minimum cardinality $T_0$-join $E_4$ in $G[V_0]$; recall that this graph is 2-edge-connected.
By Proposition~\ref{prop:frank},
$|E_4| \le \frac{1}{2} (|V_0| + \varphi_0 - 1)$.

Now  we have a connected-$T$-join with at most $|E_1|+|E_2|+|E_3|+|E_4|$ edges,
which can be bounded as follows by substituting the bounds for each of these sets,
and recalling $\varphi_0+\varphi_1+\varphi_M=\varphi(G)$:
\begin{eqnarray*}
\hspace*{0.7cm} && \hspace*{-1.7cm}
|E_1|+|E_2|+|E_3|+|E_4| \\
&\le& \textstyle
\frac{3}{2}|V_M|+\frac{1}{2}\varphi_M
\ + \ |V_0|-\mu(G,M)-1
\ + \ \frac{3}{2}|V_1|+ \frac{1}{2} \varphi_1 - (\pi - |M|)
\textstyle
\ + \ \frac{1}{2} (|V_0| + \varphi_0 - 1) \\
&=& \textstyle
\frac{3}{2}|V(G)| - 1 + |M| - \mu(G,M) + \frac{1}{2}(\varphi(G) - 1) - \pi \\
&=& \textstyle
L_{\mu}(G,M) + \frac{1}{2}L_{\varphi}(G) -\pi.
\end{eqnarray*}
\mathendproof

When the number of pendant ears is large, we will use this theorem for all the three problems.
For the complementary case
three different approaches will be needed for our three approximation algorithms.
Our first approximation algorithm deals with the connected-$T$-join problem:

\begin{theorem}\label{thm:connTjoin}
There is a $\frac{3}{2}$-approximation algorithm for the connected-$T$-join problem.
For any connected graph $G$ and $T\subseteq V(G)$ with $|T|$ even,
it finds a connected-$T$-join of cardinality at most $\frac{3}{2}\lp(G,T)$
in $O(|V(G)|^3)$ time.
\end{theorem}

\prove
We may assume that $G$ is $2$-vertex-connected (Proposition~\ref{reduction2connected}).
We construct a nice ear-decomposition that contains a maximum earmuff
for the eardrum $M$ associated with it and $T$
(using Lemma \ref{lem:new}).
Let $\pi$ be the number of pendant ears.

If $\pi\ge\frac{1}{2}\varphi(G)$,
we use Theorem~\ref{connTjoinbyearmuff} to find a connected-$T$-join of cardinality at most
$$\textstyle
L_\mu(G,M)  +   \frac{1}{2} L_{\varphi}(G)   -  \pi \ \le \ L_\mu(G,M)  +    \frac{1}{2} (|V(G)| - 1),$$
which is at most $\frac{3}{2}\lp(G,T)$ according to
Theorem~\ref{lbmuT} and the second inequality of Proposition~\ref{proplpT}.

If $\pi \le \frac{1}{2}\varphi(G)$, then
we apply Proposition~\ref{prop:connected}.
Since $\pi_2\le\pi$,
where $\pi_2$ is the number of 2-ears,
we get a connected-$T$-join of cardinality at most
$$\textstyle
\frac{3}{2} (|V(G)|-1)  + \pi -  \frac{1}{2}\varphi(G)
\ \le \ \frac{3}{2} (|V(G)|-1).$$
By  Proposition~\ref{proplpT}, this is at most $\frac{3}{2}\lp(G,T)$, and $\lp(G,T)\le\opt(G,T)$.
\endproof

\begin{figure}[ht]
\begin{center}
 \begin{tikzpicture}[scale=0.9]

  \tikzstyle{vertex}=[blue,circle,fill,minimum size=6,inner sep=0pt]
  \tikzstyle{oddvertex}=[darkgreen,fill,minimum size=5,inner sep=0pt]
  \tikzstyle{tree}=[line width=1.5]
  \tikzstyle{other}=[grey, line width=1.5]

 \node[vertex] (s)  at ( -1.5, 1) {};
 \node[vertex] (ua)  at ( -1, -0.0) {};
 \node[oddvertex] (va)  at ( -1, 2.0) {};
 \node[vertex] (ub)  at ( 0, -0.0) {};
 \node[vertex] (vb)  at ( 0, 2.0) {};
 \node[vertex] (u1)  at ( 1, -0.0) {};
 \node[vertex] (v1)  at ( 1, 2.0) {};
 \node[vertex] (u2)  at ( 2, -0.0) {};
 \node[vertex] (v2)  at ( 2, 2.0) {};
 \node[vertex] (u3)  at ( 3, -0.0) {};
 \node[vertex] (v3)  at ( 3, 2.0) {};
 \node[vertex] (u4)  at ( 4, -0.0) {};
 \node[vertex] (v4)  at ( 4, 2.0) {};
 \node[oddvertex] (u5)  at ( 5, -0.0) {};
 \node[oddvertex] (v5)  at ( 5, 2.0) {};
 \node[vertex] (u6)  at ( 6, -0.0) {};
 \node[vertex] (v6)  at ( 6, 2.0) {};
 \node[vertex] (u7)  at ( 7, -0.0) {};
 \node[vertex] (v7)  at ( 7, 2.0) {};
 \node[vertex] (u8)  at ( 8, -0.0) {};
 \node[vertex] (v8)  at ( 8, 2.0 ) {};
 \node[vertex] (u9)  at ( 9, -0.0) {};
 \node[vertex] (v9)  at ( 9, 2.0) {};
 \node[vertex] (uc)  at ( 10, -0.0) {};
 \node[vertex] (vc)  at ( 10, 2.0 ) {};
 \node[vertex] (ud)  at ( 11, -0.0) {};
 \node[oddvertex] (vd)  at ( 11, 2.0) {};
 \node[vertex] (t)  at ( 11.5, 1) {};
 \node[vertex] (a)  at ( 5, 1) {};

 \node at (-1.8,1) {\small $s$};
 \node at (11.8,1) {\small $t$};

 \draw[tree] (s) to (ua);
 \draw[other] (s) to (va);
 \draw[tree] (ua) to (ub);
 \draw[tree] (va) to (vb);
 \draw[tree] (ub) to (u1);
 \draw[tree] (vb) to (v1);
 \draw[tree] (u1) to (u2);
 \draw[tree] (v1) to (v2);
 \draw[tree] (u2) to (u3);
 \draw[tree] (v2) to (v3);
 \draw[tree] (u3) to (u4);
 \draw[tree] (v3) to (v4);
 \draw[tree] (u4) to (u5);
 \draw[tree] (v4) to (v5);
 \draw[tree] (u5) to (u6);
 \draw[tree] (v5) to (v6);
 \draw[tree] (u6) to (u7);
 \draw[tree] (v6) to (v7);
 \draw[tree] (u7) to (u8);
 \draw[tree] (v7) to (v8);
 \draw[tree] (u8) to (u9);
 \draw[tree] (v8) to (v9);
 \draw[tree] (u9) to (uc);
 \draw[tree] (v9) to (vc);
 \draw[tree] (uc) to (ud);
 \draw[tree] (vc) to (vd);
 \draw[tree] (ud) to (t);
 \draw[other] (vd) to (t);
 \draw[tree] (u5) to (a);
 \draw[tree] (v5) to (a);
 \draw[other] (ua) to (vb);
 \draw[other] (ub) to (v1);
 \draw[other] (u1) to (v2);
 \draw[other] (u2) to (v3);
 \draw[other] (u3) to (v4);
 \draw[other] (u4) to (v5);
 \draw[other] (u7) to (v6);
 \draw[other] (u8) to (v7);
 \draw[other] (u9) to (v8);
 \draw[other] (uc) to (v9);
 \draw[other] (ud) to (vc);

 \end{tikzpicture}
 \caption{\label{figctjexample}
 {\small Example showing that the computed connected-$T$-join is not necessarily
 shorter than $\frac{3}{2}$ times the optimum.
 For each $k\in\mathbb{N}$, we have a graph $G$ with $8k+5$ vertices and $12k+5$ edges.
 Two vertices are labeled $s$ and $t$; they form the set $T=\{s,t\}$.
 The figure shows the case $k=3$.
 Note that there is a Hamiltonian $s$-$t$-path, and hence $\lp(G,T)=\opt(G,T)=8k+4$.
 Also note that $\varphi(G)=2$ because $G$ is not factor-critical.
 Suppose that we choose the ear-decomposition that begins with the circuit of length $8k+4$
 and then has one pendant 2-ear (in the center).
 Then $\pi= 1= \frac{1}{2}\varphi(G)$, so we have two choices in our algorithm.
 If we use Theorem~\ref{connTjoinbyearmuff},
 then our algorithm first takes the 2-ear and then adds edges to obtain a spanning tree, e.g., the one with thick edges.
 Then there are four vertices (shown as squares) whose degrees have the wrong parity,
 and we need another $4k+2$ edges to correct the parities. So we end up with a connected-$T$-join with $12k+6$ edges.
 If we use Proposition~\ref{prop:connected} instead,
 we could also end up with $12k+6$ edges.
 }}
 \end{center}
 \end{figure}
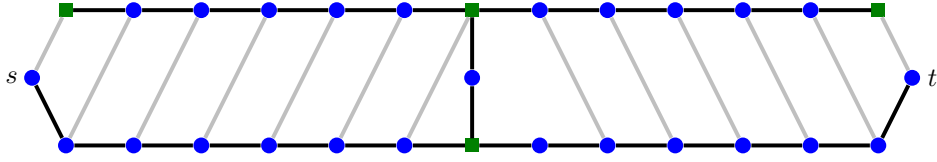

The result is tight as Figure \ref{figctjexample} shows.

\medskip

\subsection{7/5-approximation for graphic TSP}

Our algorithm for the graphic TSP will first construct a nice ear-decomposition
containing a maximum earmuff, then removes the 1-ears and computes a tour within
each block of the resulting graph. Here we distinguish two cases.
If there are many pendant ears, we get a short tour by Theorem \ref{connTjoinbyearmuff}.
If there are few pendant ears, we use the following concept of \cite{MomS11}:

\begin{definition}[Definition 3.1 of \cite{MomS11}]
\label{defpairing}
Given a connected graph $G$, a {\em removable pairing} of $G$ is a pair $(R,\Pscr)$ of sets
such that
\vspace{-6pt}
\begin{itemize}
\addtolength{\itemsep}{-6pt}
\item[-] $R\subseteq E(G)$;
\item[-] for each $P\in\Pscr$ there are three distinct edges $e,e',e''\in E(G)$ and a vertex $v\in V(G)$
with $e,e',e''\in\delta(v)$ and $P=\{e,e'\}\subseteq R$;

\item[-] for any two distinct pairs $P,P'\in\Pscr$ we have $P\cap P'=\emptyset$;

\item[-] if $S\subseteq R$ and $|S\cap P|\le 1$ for all $P\in\Pscr$,
then $(V(G),E(G)\setminus S)$ is connected.
\end{itemize}
\vspace{-5pt}
We will call the elements of $\Pscr$ simply {\em pairs}.
\end{definition}

We need the following very nice lemma and include a variant of the proof:

\begin{theorem}[Lemma 3.2 of \cite{MomS11}]
\label{ms32}
Let $G$ be a $2$-vertex-connected graph and $(R,\Pscr)$ a removable pairing.
Then $G$ has a tour of cardinality at most $\frac{4}{3}|E(G)|-\frac{2}{3}|R|$.
Moreover, such a tour can be found in $O(|V(G)|^3)$ time.
\end{theorem}

\prove
An {\em odd join} in a graph $G$ is a $T$-join in $G$ where $T$ is the set of odd degree vertices of $G$.
For any odd join $F$ in $G$ that intersects each pair $P\in\Pscr$ in at most one edge,
we construct a connected-$\emptyset$-join
from $E(G)$ by doubling the edges in $F\setminus R$ and deleting the edges in $F\cap R$.
This connected-$\emptyset$-join has $|E(G)|+c(F)$ edges, where
we define weights
$c(e)=1$ for $e\in E(G)\setminus R$ and $c(e)=-1$ for $e\in R$,
and $c(F)=\sum_{e\in F} c(e)$.

To compute an odd join of weight at most $\frac{1}{3}|E(G)| - \frac{2}{3}|R|$,
intersecting each pair at most once,
we construct an auxiliary graph $G'$ as follows.
For each pair $P=\{\{v,w\},\{v,w'\}\}\in\Pscr$ we add a vertex $v_P$ and an edge $\{v,v_P\}$ of weight zero,
and replace the two edges in $P$ by $\{v_P,w\}$ and $\{v_P,w'\}$,
keeping their weight.

$G'$ is 2-edge-connected.
Hence the vector with all components $\frac{1}{3}$ is in the convex hull
$$\bigl\{x \! \in \! [0, \! 1]^{E(G')} \! :
|F| \!- \! x(F) \! + \! x(\delta(W) \! \setminus \! F) \ge 1 \,
\mbox{ for all } W \! \subseteq \! V(G') \mbox{ and } F \! \subseteq \! \delta(W)
\mbox{ with } |\delta(W) \! \setminus \! F| \mbox{ odd}
\bigr\} $$
of incidence vectors of odd joins of $G'$, and even in the face $Q$ of
this polytope defined by $x(\delta(v_P))=1$ for all $P\in\Pscr$.
So $Q$ contains the incidence vector of an odd join $J'$ in $G'$
of weight at most $\frac{1}{3}c(E(G'))=\frac{1}{3}|E(G)| - \frac{2}{3}|R|$.
Such a $J'$ corresponds to an odd join $J$ in $G$ intersecting each pair at most once and
having weight at most $\frac{1}{3}|E(G)| - \frac{2}{3}|R|$.
To find such a $J'$ and hence such a $J$,
we add a large constant to all weights of edges incident to $v_P$ for all $P\in\Pscr$,
and find a minimum weight odd join in $G'$ with respect to these modified weights.
\endproof

We apply this in the following way:

\begin{lemma}
\label{swedishcase}
Given a 2-vertex-connected graph $G$ and an ear-decomposition
in which all ears are nontrivial,
a tour of cardinality at most
$\textstyle \frac{4}{3}(|V(G)|-1)+\frac{2}{3}\pi$
can be found in $O(|V(G)|^3)$ time,
where $\pi$ is the number of pendant ears.
\end{lemma}

\prove
In order to apply Theorem \ref{ms32}, we define a removable pairing.
For each non-pendant ear we define a pair of two edges
of the ear that share a vertex that is an endpoint of another nontrivial ear.
For each pendant ear we add any one of its edges to $R$.
This defines a removable pairing with $|R|=2k- \pi$,
where $k$ is the number of ears.
Note that $|E(G)|=|V(G)|+k-1$.
From Theorem \ref{ms32} we get then a tour of cardinality at most
$\frac{4}{3}(|V(G)|+k-1) - \frac{2}{3} (2k-\pi)=\frac{4}{3}(|V(G)|-1) + \frac{2}{3}\pi$.
\endproof

\begin{theorem}\label{thm:tsp}
There is a $\frac{7}{5}$-approximation algorithm for graphic TSP.
For any connected graph $G$ it finds a tour of cardinality at most $\frac{7}{5}\lp(G)$
in $O(|V(G)|^3)$ time.
\end{theorem}

\prove
We may assume that $G$ is $2$-vertex-connected (Proposition~\ref{reduction2connected}).
We construct a nice ear-decomposition containing a maximum earmuff
for the eardrum $M$ associated with it and $T=\emptyset$
(Lemma \ref{lem:new}).
 Define
 $\Lambda(G_,M):= \frac{2}{3}L_\mu(G,M)  +    \frac{1}{3}L_\varphi(G)$.
By Corollary~\ref{lbmu}, Theorem~\ref{lbphi} and Proposition~\ref{proplp}
we have $\Lambda(G,M)\le\lp(G)\le \opt(G)$.

 Let $G'$ be the ($2$-edge-connected, spanning) subgraph resulting from $G$ by deleting all $1$-ears.
 Note that $\varphi(G')=\varphi(G)$,
 $M$ is also the eardrum associated with the (nice) ear-decomposition without the 1-ears and $T=\emptyset$,
 and $\mu(G',M)=\mu(G,M)$.
 Therefore we also have $\Lambda(G',M)=\Lambda(G,M)$,
 and the following Claim implies the theorem.

\boldheader{Claim}
Given a graph $G'$ with a nice ear-decomposition without 1-ears,
containing a maximum earmuff for the eardrum $M$ associated with it and $T=\emptyset$,
a tour of cardinality
at most $\frac{7}{5}\Lambda(G',M)$ can be constructed in
in $O(|V(G')|^3)$ time.

\smallskip
We first prove the Claim in the case that $G'$ is 2-vertex-connected.
We use our two constructions for a tour.

If $\pi\le\frac{1}{10}\Lambda(G',M)$, then we use Lemma \ref{swedishcase} and
$|V(G')|-1 \le \Lambda(G',M)$  to obtain a tour of cardinality at most
$\frac{4}{3}\Lambda(G',M)+\frac{2}{3}\pi \le \frac{7}{5}\Lambda(G',M)$.

If $\pi \ge \frac{1}{10}\Lambda(G',M)$, then we apply
Theorem~\ref{connTjoinbyearmuff}  to $G'$, $T=\emptyset$ and $M$:
we obtain a tour of cardinality at most
$\frac{3}{2}\Lambda(G',M)-\pi \le  \frac{7}{5}\Lambda(G',M)$.

The shorter one of the two tours has cardinality at most
$\frac{7}{5}\Lambda(G',M)$.

\smallskip
To prove the Claim in the general case, we use
induction on $|V(G')|$. Suppose $v\in V(G')$ is a cut-vertex,
and $G_1$ and $G_2$ are graphs with
$G'=(V(G_1)\cup V(G_2),E(G_1)\cup E(G_2))$ and
$V(G_1)\cap V(G_2)=\{v\}$.
Then the ears $P$ with $\inn(P)\subseteq V(G_i)$ form an
ear-decomposition of $G_i$
that contains a maximum earmuff
for the eardrum $M_i$ associated with it and $T=\emptyset$
(for each $i\in\{1,2\}$). Moreover, $|M_1|+|M_2|=|M|$,
$\mu(G_1,M_1)+\mu(G_2,M_2)=\mu(G',M)$, and $|V(G_1)|+|V(G_2)|=|V(G')|+1$. Hence
\begin{eqnarray*}
L_{\mu}(G_1,M_1)+L_{\mu}(G_2,M_2)
&\!\! = \!\!&
|V(G_1)| \! - \!1+|M_1|-\mu(G_1,M_1)  +  |V(G_2)| \! - \! 1+|M_2|-\mu(G_2,M_2) \\
&\!\! = \!\!&
|V(G')| \! - \! 1+|M|-\mu(G',M) \ = \
L_{\mu}(G',M).
\end{eqnarray*}
The ear-decompositions of $G_1$ and $G_2$ contain $\varphi(G_1)$ and $\varphi(G_2)$ even ears,
respectively, and $\varphi(G_1)+\varphi(G_2)=\varphi(G')$.
Therefore we have
$$L_{\varphi}(G_1)+L_{\varphi}(G_2) \ = \
|V(G_1)|+\varphi(G_1)-1 \ + \ |V(G_2)|+\varphi(G_2)-1 \ = \
|V(G')|+\varphi(G')-1 \ = \
L_{\varphi}(G').$$

Hence
$\Lambda(G_1,M_1)+\Lambda(G_2,M_2) = \Lambda(G',M)$.
By the induction hypothesis,
a tour of cardinality
at most $\frac{7}{5}\Lambda(G_i,M_i)$ can be constructed in $G_i$ in polynomial time $(i=1,2)$.
The union of these two tours
is a tour in $G'$  of cardinality at most
$\frac{7}{5}\Lambda(G_1,M_1)+\frac{7}{5}\Lambda(G_2,M_2) = \frac{7}{5} \Lambda(G',M)$.
\endproof

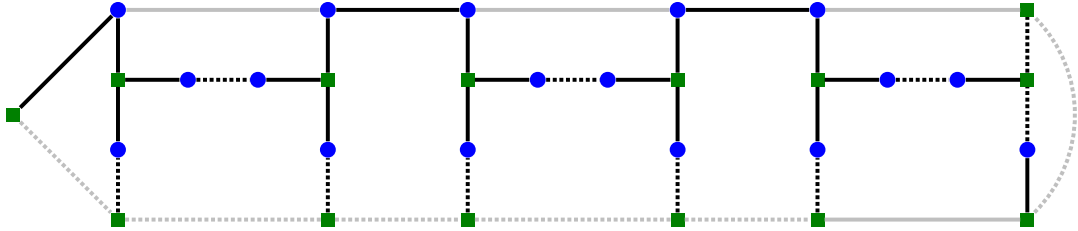
\begin{figure}[ht]
\begin{center}
 \begin{tikzpicture}[scale=0.93]

  \tikzstyle{vertex}=[blue,circle,fill,minimum size=6,inner sep=0pt]
  \tikzstyle{oddvertex}=[darkgreen,fill,minimum size=5,inner sep=0pt]
  \tikzstyle{edge}=[line width=1.5]
  \tikzstyle{redge}=[line width=1.5, densely dotted]
  \tikzstyle{other}=[grey, line width=1.5]
  \tikzstyle{rother}=[grey, line width=1.5, densely dotted]

 \node[oddvertex] (v0)  at ( 1.5, 1.5) {};
 \node[oddvertex] (v10)  at ( 3, 0) {};
 \node[vertex] (v11)  at ( 3, 1) {};
 \node[oddvertex] (v12)  at ( 3, 2) {};
 \node[vertex] (v13)  at ( 3, 3) {};
 \node[vertex] (v14)  at ( 4, 2) {};
 \node[vertex] (v15)  at ( 5, 2) {};
 \node[oddvertex] (v16)  at ( 6, 0) {};
 \node[vertex] (v17)  at ( 6, 1) {};
 \node[oddvertex] (v18)  at ( 6, 2) {};
 \node[vertex] (v19)  at ( 6, 3) {};
 \node[oddvertex] (v20)  at ( 8, 0) {};
 \node[vertex] (v21)  at ( 8, 1) {};
 \node[oddvertex] (v22)  at ( 8, 2) {};
 \node[vertex] (v23)  at ( 8, 3) {};
 \node[vertex] (v24)  at ( 9, 2) {};
 \node[vertex] (v25)  at ( 10, 2) {};
 \node[oddvertex] (v26)  at ( 11, 0) {};
 \node[vertex] (v27)  at ( 11, 1) {};
 \node[oddvertex] (v28)  at ( 11, 2) {};
 \node[vertex] (v29)  at ( 11, 3) {};
 \node[oddvertex] (v30)  at ( 13, 0) {};
 \node[vertex] (v31)  at ( 13, 1) {};
 \node[oddvertex] (v32)  at ( 13, 2) {};
 \node[vertex] (v33)  at ( 13, 3) {};
 \node[vertex] (v34)  at ( 14, 2) {};
 \node[vertex] (v35)  at ( 15, 2) {};
 \node[oddvertex] (v36)  at ( 16, 0) {};
 \node[vertex] (v37)  at ( 16, 1) {};
 \node[oddvertex] (v38)  at ( 16, 2) {};
 \node[oddvertex] (v39)  at ( 16, 3) {};

 \draw[rother] (v0) to (v10);
 \draw[edge] (v0) to (v13);
 \draw[redge] (v10) to (v11);
 \draw[edge] (v11) to (v12);
 \draw[edge] (v12) to (v13);
 \draw[rother] (v10) to (v16);
 \draw[edge] (v12) to (v14);
 \draw[redge] (v14) to (v15);
 \draw[edge] (v15) to (v18);
 \draw[other] (v13) to (v19);
 \draw[redge] (v16) to (v17);
 \draw[edge] (v17) to (v18);
 \draw[edge] (v18) to (v19);
\draw[rother] (v16) to (v20);
\draw[edge] (v19) to (v23);
 \draw[redge] (v20) to (v21);
 \draw[edge] (v21) to (v22);
 \draw[edge] (v22) to (v23);
 \draw[rother] (v20) to (v26);
 \draw[edge] (v22) to (v24);
 \draw[redge] (v24) to (v25);
 \draw[edge] (v25) to (v28);
 \draw[other] (v23) to (v29);
 \draw[redge] (v26) to (v27);
 \draw[edge] (v27) to (v28);
 \draw[edge] (v28) to (v29);
 \draw[rother] (v26) to (v30);
 \draw[edge] (v29) to (v33);
 \draw[redge] (v30) to (v31);
 \draw[edge] (v31) to (v32);
 \draw[edge] (v32) to (v33);
 \draw[other] (v30) to (v36);
 \draw[edge] (v32) to (v34);
 \draw[redge] (v34) to (v35);
 \draw[edge] (v35) to (v38);
 \draw[other] (v33) to (v39);
 \draw[edge] (v36) to (v37);
 \draw[redge] (v37) to (v38);
 \draw[redge] (v38) to (v39);
 \draw[rother] (v36) to[out=45,in=-45] (v39);

 \end{tikzpicture}
 \caption{\label{fignewtspexample}
 {\small Example showing that the computed tour is not necessarily much shorter than $\frac{7}{5}$ times the optimum.
 For each $k\in\mathbb{N}$, we have a Hamiltonian graph with $10k+1$ vertices and $13k+1$ edges.
 The figure shows the case $k=3$.
 We have $\lp(G)=\opt(G)=10k+1$ and $\varphi(G)=0$.
 Construct a nice open ear-decomposition, starting with
 $2k$ $5$-ears from left to right, each with three vertical edges,
 and then adding the $k$ horizontal pendant $3$-ears and the $1$-ear (the rightmost edge).
 Let $M$ be the eardrum associated with this ear-decomposition and $T=\emptyset$.
 We have $\Lambda(G,M)=10k$
 and $\pi=k=\frac{1}{10}\Lambda(G,M)$, so we have two choices in our algorithm.
If we use Theorem~\ref{connTjoinbyearmuff}, then our algorithm
takes first the 3-ears (they constitute a maximum earmuff). Then we could choose
the spanning tree consisting of the $10k$ black (solid and dashed) edges.
 The $4k+2$ odd degree vertices of this spanning tree are shown as squares.
 We then need another $4k$ edges
 to make all degrees even, obtaining a tour of cardinality $14k$.
 If we apply Theorem~\ref{ms32}, we delete the $1$-ear and could define the removable set $R$ as the other dotted edges.
 We have $|R|=5k$, and Theorem~\ref{ms32} provides the bound $\frac{4}{3}13k - \frac{2}{3}5k=14k$.
(In fact, if we define weights $-1$ on the dotted edges and $1$ otherwise
(cf.\ the proof of Theorem \ref{ms32}),
then the minimum weight of an odd join in $G$ that contains at most one dotted edge of each ear is $k$.
Therefore, computing such an odd join does not help here.)
 }}
 \end{center}
 \end{figure}

This result is tight as Figure \ref{fignewtspexample} shows.

\medskip

\subsection{4/3-approximation for 2ECSS}

\begin{theorem}\label{thm:2ECSS}
There is a $\frac{4}{3}$-approximation algorithm for the minimum 2-edge-connected spanning subgraph problem.
For any 2-edge-connected graph $G$ it finds a 2-edge-connected spanning subgraph with at most $\frac{4}{3}\lp(G)$ edges
in $O(|V(G)|^3)$ time.
\end{theorem}

\prove
We may assume that our graph $G$ is $2$-vertex-connected (Proposition~\ref{reduction2connected}).
We construct a nice ear-decomposition containing a maximum earmuff
for the eardrum $M$ associated with it and $T=\emptyset$
(Lemma \ref{lem:new}). Let $\pi$ denote again the number of pendant ears and
$\pi_3$  the number of (pendant) $3$-ears. We have $\pi_3\le\pi$.

\boldheader{Claim}
The number of edges in nontrivial ears is at most   $\frac{5}{4}L_\varphi(G) + \frac{1}{2}\pi$.

\smallskip
Indeed, for any ear $P$ with  $|E(P)|\ge 5$  we have $|E(P)|\le\frac{5}{4}|\inn(P)|$,
for any $2$-ear and $4$-ear we have $|E(P)|\le\frac{5}{4}|\inn(P)|+\frac{3}{4}$ (with equality for $2$-ears),
and for $3$-ears we have $|E(P)|=\frac{5}{4}|\inn(P)|+\frac{1}{2}$.
Summing up for all ears (the sum of $2$-ears and 4-ears being at most  $\varphi(G)$),
we get at most $\frac{5}{4}(|V(G)|-1)+ \frac{3}{4}\varphi(G) + \frac{1}{2}\pi_3$ edges,
implying the claim using $\pi_3\le\pi$.

\smallskip
We have now two constructions for a 2ECSS, and the better of the two satisfies the claimed bound:

If $\pi\le\frac{1}{6}\lp(G)$, then we use the Claim and
$L_\varphi (G) \le \lp(G)\le \ec(G)$ (Theorem~\ref{lbphi}, Proposition \ref{proplp}) to obtain a 2ECSS with at most
$\frac{5}{4}\lp(G)+\frac{1}{2}\pi \le \frac{4}{3}\lp(G)\le \frac{4}{3}\ec(G)$ edges.

If $\pi\ge\frac{1}{6}\lp(G)$, then we apply
Theorem~\ref{connTjoinbyearmuff} to $G$, $T=\emptyset$ and $M$:
using Theorem~\ref{lbmuT}, Theorem~\ref{lbphi} and Proposition~\ref{proplp} as before, we obtain
a tour, and hence a 2ECSS, of cardinality at most
$\frac{3}{2}\lp(G)-\pi \le  \frac{4}{3}\lp(G)\le\frac{4}{3}\ec(G)$.
\endproof

\begin{figure}[ht]
\begin{center}
 \begin{tikzpicture}[scale=0.33]

  \tikzstyle{vertex}=[blue,circle,fill,minimum size=5,inner sep=0pt]
  \tikzstyle{oddvertex}=[darkgreen,fill,minimum size=4,inner sep=0pt]
  \tikzstyle{edge}=[grey, line width=1.5]
  \tikzstyle{tree}=[line width=1.5]
  \tikzstyle{other}=[grey,line width=1.5, densely dotted]

\node[oddvertex] (v0)  at ( 2, 4.5) {};
 \node[vertex] (v06)  at ( 4, 5.25) {};
 \node[vertex] (v11)  at ( 6, 0) {};
 \node[oddvertex] (v12)  at ( 6, 3) {};
 \node[vertex] (v13)  at ( 6, 6) {};
 \node[oddvertex] (v14)  at ( 6, 9) {};
 \node[vertex] (v15)  at ( 8, 5) {};
 \node[vertex] (v16)  at ( 10, 4) {};
 \node[oddvertex] (v21)  at ( 12, 0) {};
 \node[vertex] (v22)  at ( 12, 3) {};
 \node[oddvertex] (v23)  at ( 12, 6) {};
 \node[vertex] (v24)  at ( 12, 9) {};
 \node[vertex] (v25)  at ( 14, 4) {};
 \node[vertex] (v26)  at ( 16, 5) {};
 \node[vertex] (v31)  at ( 18, 0) {};
 \node[oddvertex] (v32)  at ( 18, 3) {};
 \node[vertex] (v33)  at ( 18, 6) {};
 \node[oddvertex] (v34)  at ( 18, 9) {};
 \node[vertex] (v35)  at ( 20, 5) {};
 \node[vertex] (v36)  at ( 22, 4) {};
 \node[oddvertex] (v41)  at ( 24, 0) {};
 \node[vertex] (v42)  at ( 24, 3) {};
 \node[oddvertex] (v43)  at ( 24, 6) {};
 \node[vertex] (v44)  at ( 24, 9) {};
 \node[vertex] (v45)  at ( 26, 4) {};
 \node[vertex] (v46)  at ( 28, 5) {};
 \node[vertex] (v51)  at ( 30, 0) {};
 \node[oddvertex] (v52)  at ( 30, 3) {};
 \node[vertex] (v53)  at ( 30, 6) {};
 \node[oddvertex] (v54)  at ( 30, 9) {};

 \node[vertex] (v55)  at ( 32, 5) {};
 \node[vertex] (v56)  at ( 34, 4) {};
 \node[oddvertex] (v61)  at ( 36, 0) {};
 \node[vertex] (v62)  at ( 36, 3) {};
 \node[oddvertex] (v63)  at ( 36, 6) {};
 \node[vertex] (v64)  at ( 36, 9) {};
 \node[vertex] (v65)  at ( 38, 4) {};
 \node[vertex] (v66)  at ( 40, 5) {};
 \node[vertex] (v71)  at ( 42, 0) {};
 \node[oddvertex] (v72)  at ( 42, 3) {};
 \node[vertex] (v73)  at ( 42, 6) {};
 \node[oddvertex] (v74)  at ( 42, 9) {};
 \node[vertex] (v75)  at ( 44, 5) {};
 \node[vertex] (v76)  at ( 46, 4) {};
 \node[oddvertex] (v81)  at ( 48, 0) {};
 \node[oddvertex] (v82)  at ( 48, 3) {};
 \node[oddvertex] (v83)  at ( 48, 6) {};
 \node[vertex] (v84)  at ( 48, 9) {};

 \draw[edge] (v0) to (v11);
 \draw[edge] (v0) to (v14);
 \draw[tree] (v0) to (v06);
 \draw[tree] (v06) to (v13);
 \draw[tree] (v11) to (v12);
 \draw[edge] (v12) to (v13);
 \draw[edge] (v13) to (v14);
 \draw[tree] (v11) to (v21);
 \draw[tree] (v14) to (v24);
 \draw[tree] (v13) to (v15);
 \draw[tree] (v15) to (v16);
 \draw[tree] (v16) to (v22);
 \draw[tree] (v21) to (v22);
 \draw[tree] (v22) to (v23);
 \draw[edge] (v23) to (v24);
 \draw[tree] (v21) to (v31);
 \draw[tree] (v24) to (v34);
 \draw[tree] (v22) to (v25);
 \draw[tree] (v25) to (v26);
 \draw[tree] (v26) to (v33);
 \draw[edge] (v31) to (v32);
 \draw[tree] (v32) to (v33);
 \draw[tree] (v33) to (v34);
 \draw[tree] (v31) to (v41);
 \draw[tree] (v34) to (v44);
 \draw[tree] (v33) to (v35);
 \draw[tree] (v35) to (v36);
 \draw[tree] (v36) to (v42);
 \draw[edge] (v41) to (v42);
 \draw[edge] (v42) to (v43);
 \draw[tree] (v43) to (v44);
 \draw[edge] (v41) to (v51);
 \draw[edge] (v44) to (v54);
 \draw[tree] (v42) to (v45);
 \draw[tree] (v45) to (v46);
 \draw[tree] (v46) to (v53);
 \draw[tree] (v51) to (v52);
 \draw[edge] (v52) to (v53);
 \draw[edge] (v53) to (v54);
 \draw[tree] (v51) to (v61);
 \draw[tree] (v54) to (v64);
 \draw[tree] (v53) to (v55);
 \draw[tree] (v55) to (v56);
 \draw[tree] (v56) to (v62);
 \draw[tree] (v61) to (v62);
 \draw[tree] (v62) to (v63);
 \draw[edge] (v63) to (v64);
 \draw[tree] (v61) to (v71);
 \draw[tree] (v64) to (v74);
 \draw[tree] (v62) to (v65);
 \draw[tree] (v65) to (v66);
 \draw[tree] (v66) to (v73);
 \draw[edge] (v71) to (v72);
 \draw[tree] (v72) to (v73);
 \draw[tree] (v73) to (v74);
 \draw[tree] (v71) to (v81);
 \draw[tree] (v74) to (v84);
 \draw[tree] (v73) to (v75);
 \draw[tree] (v75) to (v76);
 \draw[tree] (v76) to (v82);
 \draw[edge] (v81) to (v82);
 \draw[edge] (v82) to (v83);
 \draw[tree] (v83) to (v84);

\draw[other] (v12) to (v16);
\draw[other] (v23) to (v25);
\draw[other] (v32) to (v36);
\draw[other] (v43) to (v45);
\draw[other] (v52) to (v56);
\draw[other] (v63) to (v65);
\draw[other] (v72) to (v76);

\draw[other] (v0) to[out=40,in=150] (v23);
\draw[other] (v32) to[out=-30,in=260] (v46);
\draw[other] (v63) to[out=150,in=80] (v46);
\draw[other] (v72) to[out=-30,in=200] (v82);

\draw[other] (v21) to[out=140,in=-20] (v12);
\draw[other] (v21) to[out=40,in=250] (v26);
\draw[other] (v21) to[out=-25,in=205]  (v41);
\draw[other] (v34) to[out=155,in=30] (v14);
\draw[other] (v34) to[out=220,in=80] (v26);
\draw[other] (v34) to[out=-40,in=160] (v43);

\draw[other] (v61) to[out=140,in=-20] (v52);
\draw[other] (v61) to[out=40,in=250] (v66);
\draw[other] (v61) to[out=-25,in=205]  (v81);
\draw[other] (v74) to[out=155,in=30] (v54);
\draw[other] (v74) to[out=220,in=80] (v66);
\draw[other] (v74) to[out=-40,in=160] (v83);

 \end{tikzpicture}
 \caption{\label{figbadexample2ec}
 {\small Example showing that the computed 2ECSS is not necessarily much shorter than $\frac{4}{3}$ times the optimum.
 For each $k\in\mathbb{N}$, we have a Hamiltonian graph with $24k$ vertices and $44k-2$ edges.
 The figure shows the case $k=2$.
 We have $\lp(G)=\opt(G)=24k$ and $\varphi(G)=1$.
 Construct a nice ear-decomposition from left to right, starting with
 $4k$ 5-ears (with black and solid grey edges),
 and finally the $4k-1$ pendant 3-ears (with solid black edges), the pendant 2-ear (on the left),
 and the 1-ears (dashed grey edges).
 Then $\pi=4k=\frac{1}{6}\lp(G)$, so we have two choices in our algorithm.
 If we use the Claim (first case of the proof of Theorem~\ref{thm:2ECSS}), we take
 all $32k-1$ edges of the $8k$ nontrivial ears.
 If we apply Theorem \ref{connTjoinbyearmuff} (note that the pendant ears
 constitute a maximum earmuff),
 we first take the pendant ears (the 2-ear and all the 3-ears),
 and then add edges to obtain a spanning tree, say the one with the $24k-1$ black edges.
 The $8k+2$ odd degree vertices are shown as squares.
We then need another $8k$ edges to make all degrees even,
and a possible choice consists of the curved dashed edges.
Then the result is a 2ECSS with $32k-1$ edges.
In fact, in both cases the computed 2ECSS is minimal.
}}
\end{center}
\end{figure}
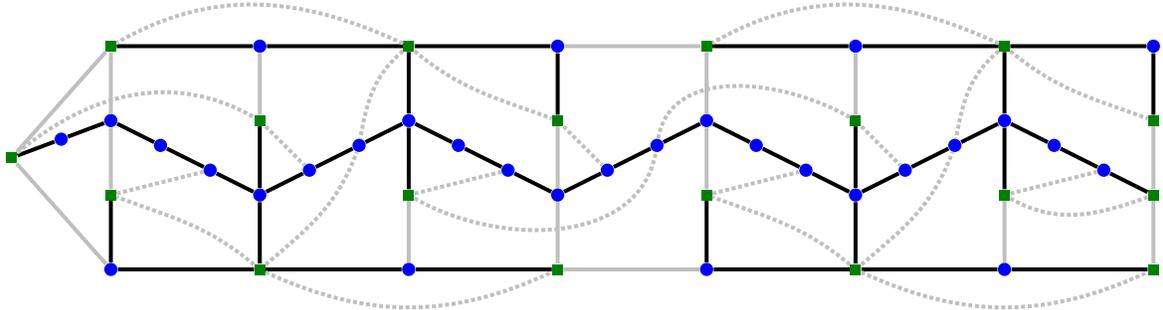

Note that the first case of the proof follows directly
from \cite{CheSS01}.
The result is tight as Figure \ref{figbadexample2ec} shows.

\section{Remarks on Integrality Ratios}

For a family $\Pscr$ of polyhedra
(say $P\subseteq\mathbb{R}^{n_P}$ for $P\in\Pscr$),
the {\em integrality ratio} of $\Pscr$ is the
supremum of the ratios
$\min\{\sum_{i=1}^{n_P}c_i x_i : x\in P\cap \mathbb{Z}^{n_P}\} / \min\{\sum_{i=1}^{n_P}c_i x_i : x\in P\}$
over all $P\in\Pscr$ and all $c\in\mathbb{R}^{n_P}$.
In this paper the objective functions are unit vectors.
By the {\em unit integrality ratio} of $\Pscr$ we mean the supremum of
$\min\{\sum_{i=1}^{n_P}x_i : x\in P\cap \mathbb{Z}^{n_P}\} / \min\{\sum_{i=1}^{n_P}x_i : x\in P\}$
over all $P\in\Pscr$.

Denote by $P(G)$ and $P(G,T)$ the polyhedra of feasible solutions of the
linear programs defining $\lp(G)$ and $\lp(G,T)$, respectively (see the Introduction).
Note that linear functions can be optimized over these polyhedra in polynomial time with the ellipsoid method:
this follows using optimization on spanning trees in polynomial time
(implying separation on the corresponding polyhedron in polynomial time),
and in addition using the max-flow-min-cut theorem, and the
algorithm of \cite{BC87} for finding a minimum weight $T$-even cut
for non-negative weight functions in polynomial time.

\begin{corollary}\label{gap:2ECSS}
For any connected graph $G$,
the integer vectors in $P(G)\cap[0,2]^{E(G)}$ correspond exactly to the $2$-edge-connected spanning subgraphs of $2G$.
The minimal integer vectors
in $P(G)$ correspond exactly to the minimal $2$-edge-connected spanning subgraphs of $2G$.
The unit integrality ratio of $\{P(G): G \mbox{ connected graph}\}$ is at most $\frac{4}{3}$.
\end{corollary}

 \prove
The first two statements are obvious, and
by Theorem~\ref{thm:2ECSS} there always exists a 2ECSS with at most $\frac{4}{3}\lp(G)$ edges.
\endproof

The integrality ratio of $\{P(G): G \mbox{ connected graph}\}$
was conjectured by \cite{CarR98} to be $\frac{4}{3}$, and Corollary \ref{gap:2ECSS} gives some support to this.
\cite{AleBE06} showed that it is at most $\frac{3}{2}$
and at least $\frac{6}{5}$ (see the example in Figure 1 of their paper).
The same example with unit weights shows that the unit integrality ratio is at least $\frac{9}{8}$.
We know no better lower bound.

\medskip
For connected-$T$-joins it does not seem useful to study the (unit) integrality ratio
of $P(G,T)$ itself, because in general not all
minimal integer vectors in $P(G,T)$ correspond to connected-$T$-joins of $G$,
not even in the case $T=\emptyset$ (indeed, $P(G,\emptyset)=P(G)$ and see Corollary \ref{gap:2ECSS}).
Therefore we intersect $P(2G,T)$ with the $T$-join polytope $Q(2G,T)$ of $2G$.
The $T$-join polytope of a connected graph $G$ is
\begin{eqnarray*}
Q(G,T) &\!\! = \!\!& \bigl\{ x\in\mathbb{R}^{E(G)}: 0\le x_e \le 1 \mbox{ for all } e\in E(G), \\
&& \hspace*{-1cm}
|F| - x(F) + x(\delta(W)\setminus F) \ge 1
\mbox{ for all } W\subseteq V(G) \mbox{ and } F\subseteq\delta(W)
\mbox{ with } |W\cap T|+ |F| \mbox{ odd} \bigr\}.
\end{eqnarray*}
We get:

\begin{corollary}\label{gap:connT}
For any connected graph $G$ and $T\subseteq V(G)$ with $|T|$ even,
the integer vectors in $P(2G,T) \cap Q(2G,T)$
are exactly the incidence vectors of connected-$T$-joins of $G$.
The unit integrality ratio of
$\{P(2G,T) \cap Q(2G,T) : G$ connected graph, $T\subseteq V(G),\, |T| \mbox{ even}\}$
is exactly $\frac{3}{2}$.
\end{corollary}

\prove
The first statement is obvious,
and by Theorem~\ref{thm:connTjoin} there always exists a connected-$T$-join of cardinality at most $\frac{3}{2}\lp(G,T)$.
This yields the upper bound.
For the lower bound, let $n\in\mathbb{N}$ and consider a circuit $G$ of length $2n$ and two vertices $s$ and $t$ at distance $n$.
The vector with all $4n$ components equal to $\frac{1}{2}$ is in $P(2G,\{s,t\}) \cap Q(2G,\{s,t\})$, but a
minimum connected-$\{s,t\}$-join has $3n$ edges.
\endproof

\begin{corollary}\label{gap:tsp}
For any connected graph $G$,
the integer vectors in $P(2G) \cap Q(2G,\emptyset)$ are exactly the incidence vectors of tours.
The unit integrality ratio of $\{P(2G) \cap Q(2G,\emptyset): G \mbox{ connected graph}\}$
is at most  $\frac{7}{5}$ and at least $\frac{4}{3}$.
\end{corollary}

\prove
The upper bound follows from Theorem~\ref{thm:tsp}.

To prove the lower bound, we consider the standard example:
let $k\in\mathbb{N}$ and define a graph $G$ as the union of three internally
vertex-disjoint paths of length $k$, all with the same endpoints.
Then the vector $x\in\mathbb{R}^{E(2G)}$ with all components $\frac{1}{2}$
is in $P(2G) \cap Q(2G,\emptyset)$ and has
$x(E(2G))=|E(G)|=3k$, but $\opt(G)=4k$.
\endproof

For a connected graph $G$, let $(\bar G,\bar c)$ again denote the metric closure of $G$, and let
$S(\bar G):=\bigl\{x\in[0,1]^{E(\bar G)}\cap P(\bar G):
x(\delta(v))= 2 \mbox{ for all } v\in V(\bar G)\bigr\}$ be the subtour polytope of $\bar G$.
Since $\lp(G) = \min\bigl\{\sum_{e\in E(\bar G)}\bar c(e) x_e : x\in P(\bar G) \bigr\}
\le \min\bigl\{\sum_{e\in E(\bar G)}\bar c(e) x_e : x\in S(\bar G) \bigr\}$,
Corollary~\ref{gap:tsp} implies an upper bound of $\frac{7}{5}$ of the integrality ratio of
the subtour polytope restricted to such ``graphic'' weight functions $\bar c$.
No better bound than $\frac{3}{2}$ (which is due to \cite{Wol80}) is known for general metric weight functions.

For general connected-$T$-joins we
have $\lp(G,T) = \min\bigl\{\sum_{e\in E(\bar G)}\bar c(e) x_e : x\in P(\bar G,T) \bigr\}$ and the ratio $\frac{3}{2}$.
Note that $P(G,\{s,t\})$ is different from the relaxation
for which \cite{AKS12} proved ratios between $1.61$ and $1.62$.

\medskip
We conclude with a remark concerning the relation between
the 2ECSS problem and the graphic TSP:

\begin{theorem}
Let $\rho\ge 1$.
If there is a $\rho$-approximation algorithm for the 2ECSS problem,
then there is a $\frac{2}{3}(\rho+1)$-approximation algorithm for the graphic TSP.
If the unit integrality ratio of $\{P(G): G \mbox{ connected graph}\}$ is $\rho$, then
the unit integrality ratio of $\{P(2G)\cap Q(2G,\emptyset): G \mbox{ connected graph}\}$
is at most  $\frac{2}{3}(\rho+1)$.
\end{theorem}

\prove
Let $G$ be a connected graph, and let $G'$ be a 2ECSS of $2G$.

\smallskip
\boldheader{Claim} $G$ has a tour of cardinality at most $\frac{2}{3}(|E(G')|+|V(G)|-1)$.

\smallskip
We prove the Claim by induction on the number of vertices.
If $G'$ is 2-vertex-connected, find any ear-decomposition of $G'$, and
define a removable pairing $(R,\Pscr)$ by including one edge of each ear in $R$
and setting $\Pscr=\emptyset$.
We have $|R|=|E(G')|-|V(G')|+1$.
By Theorem \ref{ms32} we get a tour of cardinality at most
$\frac{4}{3}|E(G')|-\frac{2}{3}|R|=\frac{2}{3}(|E(G')|+|V(G')|-1)$
as required.
If $G'$ has a cut vertex $v$, we apply the induction hypothesis to two graphs
that share only $v$ and whose union is $G'$ (as in the proof of Theorem \ref{thm:tsp}).
The Claim follows.

\smallskip
The proof is finished easily using the Claim and Proposition \ref{proplp} as follows.
If $G'$ has at most $\rho\,\ec(G)$ edges, then our tour has cardinality at most
$\frac{2}{3}(\rho\,\ec(G)+\opt(G))\le \frac{2}{3}(\rho+1)\opt(G)$.
If $G'$ has at most $\rho\,\lp(G)$ edges, then our tour has cardinality at most
$\frac{2}{3}(\rho\,\lp(G)+\lp(G))= \frac{2}{3}(\rho+1)\lp(G)$.
\endproof

This strengthens a result of \cite{MonMP90} who gave the bound $\frac{4}{3}\rho$ instead of $\frac{2}{3}(\rho+1)$.
We conclude that any $\rho$-approximate 2ECSS with $\rho<\frac{11}{10}$ leads to a
tour with less than $\frac{7}{5}\opt(G)$ edges.

\subsection*{Acknowledgment}
Many thanks to Attila Bern\'ath, Joseph Cheriyan, Satoru Iwata, Neil Olver, Zolt\'an Szigeti, Kenjiro Takazawa
and L\'aszl\'o V\'egh
for  their careful reading and suggestions,
and in particular to
Anke van Zuylen and Frans Schalekamp for in addition pointing out flaws in a preliminary version of this paper.

\small
\newcommand{\bib}[3]{\bibitem[\protect\citeauthoryear{#1}{#2}]{#3}}

\end{document}